\documentstyle[12pt,epsf]{article}

\renewcommand{\rm}{\mathrm}   

    \newcommand{\tdot}{\!\cdot\!}   \newcommand{\nsdot}{n^*\!\!\cdot\!}

    \newcommand{\dotg}{\!\!\!/}     \newcommand{\pslash}{p_{\!}\!\!/}
    \newcommand{\dd}{{\rm d}}       \newcommand{\td}{\!{\rm d}}
    \newcommand{\re}{{\rm Re}\,}
                                           
\newcommand{\eps}{{\textstyle\epsilon}}    \newcommand{\Eq}{Eq.\,}
\newcommand{\half}{{\textstyle\frac12}}    \newcommand{\Eqs}{Eqs.\,}

\def\RMP{{\em Rev. Mod. Phys.}}   \def\JMP{{\em J. Math. Phys.}}
\def\PRD{{\em Phys. Rev.} D}      \def\NPB{{\em Nucl. Phys.} B}

\begin{document}

\pagestyle{empty}
\begin{flushright}  CERN-TH/99-201 \\ July 1999  \end{flushright}

\vspace*{3mm}
\begin{center}
  {\bf Two-Loop Quark Self-Energy in a New Formalism} \\
  {\bf (II) Renormalization of the Quark Propagator} \\
  {\bf in the Light-Cone Gauge} \\

  \vspace*{0.5cm}
  {\bf George Leibbrandt} \footnote{\tt e-mail addresses:
  gleibbra@msnet.mathstat.uoguelph.ca; mthgeorg@mail.cern.ch}
  \\ Theoretical Physics Division, CERN, CH - 1211 Geneva 23\\
  and\\ Department of Mathematics and Statistics,
  University of Guelph,\\ Guelph, Ontario, Canada, N1G 2W1\\

  \vspace*{0.2cm} and\\
  \vspace*{0.2cm}
  {\bf Jimmy D. Williams} \footnote{\tt e-mail address:
  jimmydw@eudoramail.com} \\ Department of Physics,
  University of Guelph, \\ Guelph, Ontario, Canada N1G 2W1\\
  \vspace*{0.5cm}

  ABSTRACT   \end{center}

The complete two-loop correction to the quark propagator, consisting
of the spider, rainbow, gluon bubble and quark bubble diagrams, is
evaluated in the noncovariant light-cone gauge (lcg), $n \tdot A^a(x)
= 0$, $n^2 = 0$.  (The overlapping self-energy diagram had already
been computed.)  The chief technical tools include the powerful {\em
matrix integration technique}, the $n^*_\mu$-prescription for the
spurious poles of $(q\tdot n)^{-1}$, and the detailed analysis of the
boundary singularities in five- and six-dimensional parameter space.
It is shown that the total divergent contribution to the two-loop
correction $\Sigma_2$ contains both covariant and noncovariant
components, and is a {\it local} function of the external momentum
$p$, even off the mass-shell, as all {\it nonlocal} divergent terms
cancel exactly.  Consequently, both the quark mass and field
renormalizations are local.  The structure of $\Sigma_2$ implies a
quark mass counterterm of the form $\delta m (lcg) =
m\widetilde\alpha_s C_F(3+\widetilde\alpha_sW) + {\rm O}
(\widetilde\alpha_s^3)$, $\widetilde\alpha_s \equiv g^2\Gamma(\eps)
(4\pi)^{\eps -2}$, with $W$ depending only on the dimensional regulator
$\eps$, and on the numbers of colors and flavors.  It turns out that
$\delta m(lcg)$ is identical to the mass counterterm in the general
linear covariant gauge.  Our results are in agreement with the
Bassetto-Dalbosco-Soldati renormalization scheme.

\vspace{3mm}


\vspace*{3mm}   \vfill\eject


\setcounter{page}{1}  \pagestyle{plain}  

\section{Introduction}

A quarter of a century ago the light-cone gauge was a gauge ``to
fortune and to fame unknown".  It was regarded as a freakish member of
the family of axial-type gauges that existed more by accident than by
inventive planning \cite{L1}.  Today, the light-cone gauge enjoys
respect and a privileged status among noncovariant gauges for the
following two reasons:  first, computations in the light-cone gauge
have proved to be meaningful, both at one and two loops.  Second, the
renormalizability of light-cone QCD, as demonstrated in this article,
has finally been established at two-loop order by explicit computation.
Specifically, we shall discuss here in some detail the renormalization
of the quark propagator to two loops in the light-cone gauge
$n\tdot A^a=0, n^2=0$, where $A^a_\mu$ denotes the gauge field and
$n_\mu = (n_0, \vec n)$ is an arbitrary, but fixed, four-vector,
$\mu = 0,1,2,3$ \cite{L5}.  The diagrams for this process are depicted
in Figure 1. We note that the results for the one-loop quark
self-energy function (Fig.\,1f) and for the overlapping self-energy
function (Fig.\,1b) have already been reported in the literature
\cite{L3,me}.  These two diagrams are included here for completeness.

\begin{figure}[h]
\epsffile{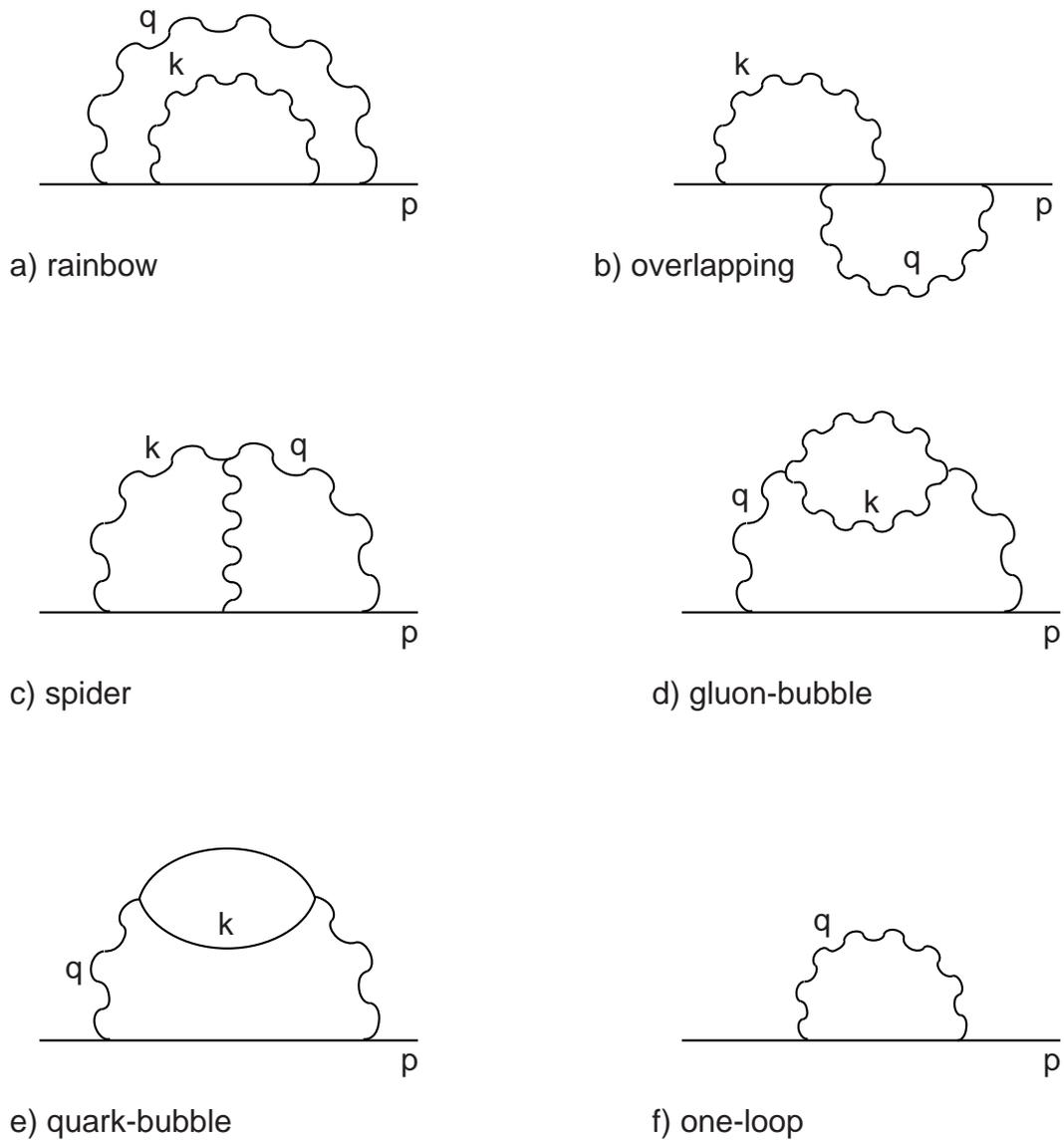}
\caption[]{One- and two-loop quark self-energy diagrams for QCD.  Wavy
lines denote gluons, straight lines denote quarks.  $p, q,$ and $k$ are
four-momenta.}
\end{figure}

Whereas computation of Fig.\,1f was straightforward, at least in
hindsight, evaluation of the overlapping quark self-energy
\cite{me,me2} required the introduction of a new procedure, called
the {\it matrix integration technique}.  We recall that in this
procedure, the two momentum integrals of $\int d^{2\omega} q \int
d^{2\omega} k f(q,k)$, where $2\omega$ denotes the complex
dimensionality of space-time, are integrated over
$4\omega$-dimensional space in a {\it single operation}.  We
further recall that the biggest advantage of the matrix method is
the ability to execute the momentum integrations {\it exactly} and
in {\it closed form}.  With the momentum integrals conveniently
``out of the way", we can then concentrate on the wide variety of
new parameter singularities which is so characteristic of
noncovariant-gauge multi-loop integrals. It turns out that the
matrix method enables us to handle these new parameter
singularities in a consistent and unambiguous manner. By
comparison, multi-loop integrals
with fewer and less severe parameter singularities
may, in general, be evaluated by means of the {\it nested method}
\cite{me,me2,noncovt,HB}. In this traditional approach, the
$2\omega$-momentum integrations are carried out sequentially. 

We finally observe that the matrix integration technique works for
covariant and noncovariant gauges alike, and regardless whether
the integrals are massive or massless. We shall have occasion
later in this article to highlight specific technical features of
this powerful technique.

\clearpage

It is well known \cite{tH,co} that in {\it covariant} gauges,
QCD may be renormalized by a redefinition of the masses, coupling
constants, and field normalizations. However, for noncovariant gauges,
such as the light-cone gauge, the situation is more complicated:  in
addition to the same types of counterterms that arise in covariant
gauges, for light-cone QCD we also require $n_\mu$-dependent
counterterms. These {\it noncovariant counterterms} are different from
any of the terms in the original Lagrangian density. On the other hand,
and in contrast to the covariant case, no physical measurements are
needed to fix the finite parts of these new noncovariant counterterms.
Instead, the finite parts are constrained by the requirement that all
physical observables be Lorentz-covariant. In a sense, the presumption
of covariance of the observables constitutes an infinite set of
physical constraints on the noncovariant counterterms.

In 1987, Bassetto, Dalbosco, and Soldati (BDS) \cite{B2,B1}
proposed a renormalization scheme for light-cone QCD in which the
particular types of noncovariant counterterms -- permitted by gauge
symmetry and Lorentz symmetry -- are absorbed into the original
Lagrangian density by means of {\it noncovariant} renormalizations of
the quark and gluon fields.  Unlike the covariant-gauge case,
where the renormalization factors are scalars, in the BDS scheme the
renormalization factors are {\it matrices}.  The result is that {\it
the different spin-components of the quark and gluon fields are
renormalized differently}. The present two-loop
calculations are a clear vindication of the BDS renormalization scheme.

The plan of paper (II) is as follows. In Section 2 we derive the
integrands of the light-cone integrals for the various diagrams of
Figure 1, and then break them down into approximately 80 simpler
integrands. In Section 3 we illustrate, by means of
examples, methods for handling the various integrations. The explicit
values for the divergent parts of all integrals are tabulated in the
Appendix. These results are analytic and allow for general quark
masses. The counterterms needed for the renormalization of the quark
self-energy to two loops in the light-cone gauge are derived in
Section 4. It turns out that all required counterterms are {\it local}
(that is, polynomial in the external momentum), and that their
coefficients satisfy the relationships implied by the BDS
renormalization scheme. The paper concludes in
Section 5 with a summary of our results and their significance.

%
\bigskip
\section{Derivation of the Integrals}

The Lagrangian density for light-cone QCD reads
\begin{equation} \label{eq:L}
  {\cal L} \,=\ -{\textstyle\frac14} F^a_{\mu\nu} F^{a\mu\nu} +
     \overline\psi_\alpha (i\partial\dotg-m) \psi_\alpha +
     gT^a_{\alpha\beta} \overline\psi_\alpha A\dotg^a \psi_\beta
     \,+{\cal L}_{\rm fix}+{\cal L}_{\rm gh}+{\cal L}_{\rm ct}\,,\quad
\end{equation}
where external source terms and quark flavor indices have been
suppressed, and

\smallskip
$F^a_{\mu\nu} =\,\partial_\mu A^a_\nu - \partial_\nu A^a_\mu +
      gf^{abc} A^b_\mu A^c_\nu\,,\qquad \mu,\nu = 0\,.\,.\,3, \qquad
                   \partial_\mu \equiv \partial/\partial x^\mu, $

$A^a_\mu(x) =\,$ gluon fields,$\qquad
      a,b,c =\,$ gluon indices$\,= 1\,.\,.\,8\ $ for SU(3),

$\psi_\alpha(x) =\,$ quark fields,$\qquad
      \alpha,\beta =\,$ color indices$\,= 1,2,3, \qquad
      \overline\psi \equiv \psi^\dagger \gamma^0, $

$A\dotg             \equiv \gamma\tdot A,      \quad
      \partial\dotg \equiv \gamma\tdot\partial, \quad $etc.$, \quad
      \gamma^\mu =\,$ Dirac matrices,$ \quad (\gamma^0)^2=1,$

$m =\,$ quark rest mass,$\quad g =\,$ coupling constant,

$T^a_{\alpha\beta} =\,$ generators of SU(3), normalized such that
$[T^a,T^b]=if^{abc}T^c$,

$f^{abc} =\,$ antisymmetric structure constants of SU(3) (with
            $f^{abc}f^{bcd} = 3\delta^{ad}$),

${\cal L}_{\rm fix} =\,$ gauge-fixing term $\,=\, -(n\tdot A)^2/
(2\lambda),\ $with $\lambda$ to be taken to zero later,

${\cal L}_{\rm gh} =\,$ ghost terms $\,=\, -\overline\eta^a
(\delta^{ab} n\tdot\partial - gf^{abc}n\tdot A^c)\eta^b,
\ \eta^a =\,$ ghost fields, and

${\cal L}_{\rm ct} =\,$ counterterms.

\smallskip
\noindent
Repeated indices imply summation, and we use a $\,+,-,-,-\,$ metric for
Minkowski space.

  \subsection{The Feynman Rules}

The Lagrangian (\ref{eq:L}) leads to the following Feynman rules for
light-cone QCD \cite{B1,L2,su}:

\smallskip
Quark propagator:
\begin{equation} \label{eq:quark}
   i\delta_{\alpha\beta} S(p),\quad {\rm with} \qquad  S(p) \equiv
   \frac {\pslash +m}{p^2-m^2+i\theta}\,,\qquad \theta > 0,
\end{equation}
where $p_\mu$ is the quark 4-momentum, $\alpha$ and $\beta$ are color
indices as in (\ref{eq:L}), and the $i\theta$ term is Feynman's
prescription for avoiding a singularity when $p^2=m^2$.  (We let
$\theta$ go to zero after  Wick rotation.)

\smallskip
Gluon propagator: $\quad i\delta^{ab} G_{\mu\nu}(q),\ \ $ with
\begin{equation} \label{eq:prop}
   G_{\mu\nu}(q)    \equiv \frac{-1}{q^2+i\theta}
    \left( g_{\mu\nu} - \frac{ n_\mu q_\nu + q_\mu n_\nu }{ n\tdot q }
    + \frac{ \lambda q^2 q_\mu q_\nu }{ (n\tdot q)^2} \right),
    \qquad \theta > 0,
\end{equation}
where $q_\mu$ is the gluon 4-momentum.  The gauge-fixing parameter
$\lambda$ is now
taken to zero, causing the third term in parentheses to drop out.

\bigskip
Quark-quark-gluon vertex factor: $\quad ig\gamma^\mu
T^a_{\alpha\beta}$.

\bigskip
3-gluon vertex factor: $\quad gf^{abc} \big[ (p-q)^\rho g^{\mu\nu} +
(q-k)^\mu g^{\nu\rho} + (k-p)^\nu g^{\mu\rho} \big], \quad $
where $p,q,k$ are the incoming 4-momenta of the attached gluons.

\bigskip
4-gluon vertex factor: $\ \ -ig^2 \big[
  f^{abe}f^{cde} (g^{\mu\rho}g^{\nu\sigma}-g^{\mu\sigma}g^{\nu\rho}) +
  f^{ace}f^{bde} (g^{\mu\nu}g^{\rho\sigma}-g^{\mu\sigma}g^{\nu\rho}) +
  f^{ade}f^{cbe} (g^{\mu\rho}g^{\nu\sigma}-g^{\mu\nu}g^{\rho\sigma})
  \big]  $.

\bigskip
Ghost propagator: $\quad \delta^{ab}/(n\tdot k),\quad$ where $k_\mu$
is the ghost 4-momentum.

\bigskip
Ghost-ghost-gluon vertex factor: $\ \ \,-igf^{abc}n^\mu,\quad$ where
$\mu$ and $a$ match indices of the attached gluon, while $b$ and $c$
match indices of the attached ghosts.  Note that ghosts ``decouple''
in the light-cone gauge, because the ghost-ghost-gluon vertex factor
is orthogonal to the gluon propagator (\ref{eq:prop}) when $\lambda=0$
\cite{fr,L1}.

\bigskip
Counterterm vertex factors: $\quad$ to be determined in Section 4.

\bigskip
To construct the dimensionally regularized Green function for a
diagram, we first impose conservation of momentum at each vertex, so
that, apart from external momenta, there is only one independent
momentum per loop. (In Figure 1, these momenta are denoted by $q$ and
$k$.)  We then form the product of the propagators and vertex factors
for all internal lines and vertices, divide by $(2\pi)^{2\omega}$ for
each loop, and integrate the resulting expression over the
$2\omega$-dimensional space of each loop momentum.
Finally, we multiply the integral by $-1$ for each internal quark loop,
divide by $r!$ for each pair of vertices connected by $r$ gluon lines,
and also divide by the number of permutations of the vertices which
leave the diagram invariant (for fixed external lines) \cite{B1,ry}.
In Figure 1, only diagram e (quark-bubble) has an internal quark loop,
only diagram d (gluon-bubble) has a pair of vertices connected by more
than one gluon line, and {\em no} diagram is invariant under a
non-trivial permutation of its vertices.

The procedure just described  gives only the ``amputated''
Green function, since it includes no factors for the external lines of
the diagram.

  \subsection{The Integrals and the $n_\mu^*$-Prescription}

To illustrate the application of the above Feynman rules, we may
immediately write down the amputated Green function for the one-loop
diagram of Figure 1f:
\begin{equation} \label{eq:a}  \delta_{\alpha\kappa} \Sigma_1(p)\ =
\ \ \frac{ -g^2T^a_{\alpha\beta}T^a_{\beta\kappa} }{ (2\pi)^{2\omega} }
  \int_M \gamma^\mu S(p-q) \gamma^\nu \,G_{\mu\nu}(q)\, \dd^{2\omega}q,
\ \end{equation}
where $\int_M$ denotes integration over Minkowski space, and $S$ and
$G$ are the propagators defined in equations (\ref{eq:quark}) and
(\ref{eq:prop}) (with $\lambda=0$, and $m$ equal to the rest mass of
the external quark).  To avoid a singularity in $G$ when
$n\tdot q=0$ (in this and other integrals), we shall use the
$n_\mu^*$-prescription (ML-prescription) \cite{ma,L5}, in which
\begin{equation} \label{eq:ML}
   \frac 1{n\tdot q} \ \to\ \lim_{\theta \to 0}\,\frac{\nsdot q}
       {\nsdot q\,n\tdot q +i\theta} \quad =\ \lim_{\theta \to 0}
        \,\frac{2 \nsdot q}{\nsdot n\,q_\|^2 +2i\theta}\,,\qquad
        \theta > 0,
\end{equation}
where $n_\mu^*$ is a new light-like 4-vector
($n^{*2}=0$), with $\nsdot n>0$ in Minkowski space, and
\begin{equation}
  q_{\|\mu}\,\equiv\,\frac{\nsdot q\,n_\mu+n\tdot q\,n^*_\mu}{\nsdot n}
  \,,\qquad\quad q_{\perp\mu}\,\equiv\,q_\mu-q_{\|\mu}    \label{eq:pp}
\end{equation}
for any 4-vector $q_\mu$.  The same prescription applies to
$(n\tdot k)^{-1}$, with the same $n^*_\mu$, but with $k$ in place of
$q$.  Prescription  (\ref{eq:ML}) was subsequently recovered in the
context of canonical quantization by Bassetto et al.\,\cite{BDLS}.
Unlike the old ``principal value'' prescription, the
$n^*_\mu$-prescription is consistent with both Wick rotation and
power-counting \cite{L1,HB,B1,L2}.

\smallskip
The amputated Green functions for the five two-loop diagrams of
Figure 1 possess the following structure:

\smallskip \noindent     Rainbow diagram:  $\quad ig^4
   T^a_{\alpha\beta} T^b_{\beta\gamma} T^b_{\gamma\delta}
   T^a_{\delta\kappa} (2\pi)^{-4\omega} I_{\rm a},\ \ $ with
\[
   I_{\rm a}\ \equiv\ \int_M \td^{2\omega}q \int_M \td^{2\omega}k \,
  \gamma^\mu S(p-q) \gamma^\nu S(p-q-k) \gamma^\rho S(p-q)
  \gamma^\sigma \,G_{\mu\sigma}(q)\,G_{\nu\rho}(k).          \]

\smallskip \noindent     Overlapping diagram:  $\quad ig^4
   T^a_{\alpha\beta} T^b_{\beta\gamma} T^a_{\gamma\delta}
   T^b_{\delta\kappa} (2\pi)^{-4\omega}I_{\rm b},\ \ $ with
\[
   I_{\rm b}\ \equiv\ \int_M \td^{2\omega}q \int_M \td^{2\omega}k \,
  \gamma^\mu S(p-q) \gamma^\nu S(p-q-k) \gamma^\rho S(p-k)
  \gamma^\sigma \,G_{\mu\rho}(q)\,G_{\nu\sigma}(k).          \]

\smallskip \noindent     Spider diagram:  $\quad g^4 f^{abc}
   T^a_{\alpha\beta} T^b_{\beta\gamma} T^c_{\gamma\kappa}
   (2\pi)^{-4\omega} I_{\rm c},\ \ $ with
\begin{eqnarray*} I_{\rm c}
     \!& \equiv &\! \int_M \td^{2\omega}q \int_M \td^{2\omega}k
     \,\gamma^\sigma S(p-q) \gamma^\eta S(p-k) \gamma^\tau \,
     G_{\mu\sigma}(q)\, G_{\nu\eta}(k-q)\, G_{\rho\tau}(k) \quad
    \\ && \qquad\qquad\qquad \cdot \big[ (k-2q)^\rho g^{\mu\nu} +
 (q-2k)^\mu g^{\nu\rho} + (q+k)^\nu g^{\mu\rho} \big]. \end{eqnarray*}

\smallskip \noindent     Gluon-bubble:  $\quad \frac12 ig^4
   f^{abc}f^{bcd} T^a_{\alpha\beta} T^d_{\beta\kappa}
   (2\pi)^{-4\omega} I_{\rm d},\ \ $ with
\begin{eqnarray*} I_{\rm d}
    \!& \equiv &\! \int_M \td^{2\omega}q \int_M \td^{2\omega}k
    \,\gamma^\phi S(p-q) \gamma^\xi\, G_{\mu\phi}(q)\,G_{\nu\eta}(k-q)
    \, G_{\rho\tau}(k)\, G_{\sigma\xi}(q)        \\
   && \qquad\qquad\quad     \cdot \big[ (k-2q)^\rho g^{\mu\nu} +
   (q-2k)^\mu g^{\nu\rho} + (q+k)^\nu g^{\mu\rho} \big]   \\
   && \qquad\qquad\quad     \cdot \big[ (k-2q)^\tau g^{\sigma\eta} +
   (q-2k)^\sigma g^{\eta\tau} + (q+k)^\eta g^{\sigma\tau} \big].
\end{eqnarray*}

\smallskip \noindent     Quark-bubble:  $\quad -ig^4 T^a_{\alpha\beta}
   T^a_{\gamma\delta} T^b_{\delta\gamma} T^b_{\beta\kappa}
   (2\pi)^{-4\omega} I_{\rm e},\ \ $ with $\ \ I_{\rm e}\ \equiv $
\begin{equation} \label{eq:f}
  \int_M \td^{2\omega}q \int_M \td^{2\omega}k \,\gamma^\mu S(p-q)
  \gamma^\nu G_{\mu\sigma}(q)             \frac{ {\rm Trace}
  [ \gamma^\sigma (k\dotg+m_L) \gamma^\rho (k\dotg-q\dotg+m_L) ] }
  { [ k^2-m_L^2+i\theta ][ (k-q)^2-m_L^2+i\theta ] } G_{\rho\nu}(q),
\ \ \end{equation} where $m_L$
is the rest mass of the quark in the {\it inner} loop of Figure 1e.

The first step in the evaluation of integrals $I_{\rm a}$ to
$I_{\rm d}$ is to substitute for $S$ and $G$ from equations
(\ref{eq:quark}) and (\ref{eq:prop}), and expand the numerator of each
integrand into a sum of products.  In order to avoid terms whose
integrals diverge as $q\to 0$, the quark-bubble integral $I_{\rm e}$
will be handled differently from the other integrals, as discussed in
Section 3.2.

\smallskip
Before integrating, we simplify  the integrals by applying {\em Wick
rotations}.   This procedure is valid because both Feynman's $i\theta$
prescription as well as the $n^*_\mu$-prescription lead to poles in
the second and fourth quadrants only. In this way, the integral
(\ref{eq:a}), for instance, becomes
\[
  \delta_{\alpha\kappa} \Sigma_1(p)\ =\, \ \frac{ -ig^2
    T^a_{\alpha\beta}T^a_{\beta\kappa} }{ (2\pi)^{2\omega} } \int_E
    \td^{2\omega}q \frac{ \gamma^\mu (\pslash-q\dotg-m) \gamma^\nu }
                        { [(p-q)^2+m^2]\,q^2 } \left( \delta_{\mu\nu}
  - \frac{ n_\mu q_\nu + q_\mu n_\nu }{ n\tdot q } \right),
\]
where $\,\mu,\nu=1\dots 4,\ q_0=iq_4,\ p_0=ip_4,\ n_0=in_4,
\ \gamma^0=i\gamma^4,\,$ and $\int_E$ denotes integration over the
Euclidean space spanned by $q_1\dots q_4$.

  \subsection{Expansion, Reduction, Transformation, and Tadpoles}

{} From the Feynman rules, we see that Green functions in
general are integrals of rational functions of the loop momenta.
For integrals $I_{\rm a}$ to $I_{\rm d}$ above, it would appear that
some terms in the numerators of the integrands could have degrees as
high as 11 in $q$ and $k$ together (after up to four applications of
the $n^*_\mu$-prescription (\ref{eq:ML})).
Fortunately, however, we may reduce the maximum degree to five, and in
most cases to three, by means of cancellations between terms within
each integral, and between numerator and denominator factors within
many of the terms.  Because of the large number of terms involved, some
computer assistance is advantageous.

Expanding the numerator of each integrand into a sum of products,
and then following the procedure outlined in paper (I), we obtain
a list of well over 100 distinct terms to be integrated. Fortunately,
we can shorten this list  still further by applying transformations,
such as $k\to q-k,\ q\to q+k,$ and/or $q\leftrightarrow k$, to selected
terms.  The transformation $q\to q+k$, for instance, yields
 \[
  \frac{ n\dotg }{ (n\tdot q-n\tdot k) (q-k)^2 [(p-q)^2+m^2] k^2 }
  \ \to\ \frac{ n\dotg }{ n\tdot q\,q^2 [(p-q-k)^2+m^2] k^2}\,,
\]
where the expression on the right just happens to be already on the
list of terms to be integrated.  Notice that the factor
$(n\tdot q-n\tdot k)$ can thus be eliminated  from all denominators.

Having carried out these various cancellations and transformations,
we find that some of
the integrands factor into separate $q$ and $k$ dependent parts. In
some cases, one of these two one-loop integrals vanishes, since its
integrand is antisymmetric under $q\to -q$ or
$k\to -k$. In many other cases, one of the one-loop integrals
corresponds to  a {\em massless tadpole}, and likewise vanishes in the
context of  dimensional regularization \cite{L6,L4}.

   \subsection{Power Counting}

For renormalization, we require only the divergent parts of
integrals $I_{\rm a}$ to $I_{\rm e}$, so we shall drop all terms whose
integrals can be shown by power counting to converge when $\omega=2$.
For a {\em covariant} gauge, Weinberg's theorem \cite{we,ry} tells us
that a Feynman integral is UV-convergent if its integrand, including
the measure $d^4q\,d^4k\dots$, is of negative degree with respect to
every non-empty subset of the loop momenta $q,k,\dots$.  We have the
same rule for the light-cone gauge,  except that we must {\em also}
consider subsets that include only the ``transverse'' part -- the
part orthogonal to $n$ -- of one or more of the loop momenta \cite{B1}.

To illustrate power counting, consider the Euclidean-space integral
\begin{equation}
  \int\int \frac{ n\dotg\,n\tdot p\, \dd^{2\omega}q \,\dd^{2\omega}k }
  { n\tdot q [(p-q)^2+m^2] (q-k)^2 (n\tdot q-n\tdot k) [(p-k)^2+m^2] }
\,,
\end{equation}
arising from the spider diagram (Figure 1c).  When $\omega=2$, the
integrand has degree $-2$ in $q$, $-1$ in $k$, and 0 in $q$ and $k$
{\em combined}.  Hence, by the rule given above, this integral
{\em may} be divergent.  As it turns out, however, this particular
integral is actually {\em con}vergent, because the leading-order part
of the integrand for large $|q|$ and $|k|$ is antisymmetric under
$q_3\leftrightarrow q_4,k_3\leftrightarrow k_4$ (we have taken
$\,n_1=n_2=n^*_1=n^*_2=0\,$ for simplicity, so that $\,q_{\|\mu}=
(0,0,q_3,q_4)\,$ and $\,k_{\|\mu}=(0,0,k_3,k_4)$).  There are other
such ``borderline'' integrals which are convergent for the same reason.

\bigskip
In the Appendix we have summarized the divergent terms which remain to
be integrated, for each of the  integrals $I_a$ to $I_e$. We have also
listed there the various integrated divergent parts for each individual
term. In the next section we shall demonstrate how these results were
obtained.

%
\bigskip
\section{Integration Methods}

Several different approaches to the evaluation of multi-loop Feynman
integrals appear in the literature \cite{noncovt,HB,im,TF,sm}; they
will not be reviewed here.  In most methods, one begins by writing the
denominator of the integrand as an integral, using a {\em
parametrization formula}.  We shall use the formula known as {\em
Schwinger's representation}:
\begin{equation} \label{eq:param}
  \frac1{F_1^{\rm u} F_2^{\rm v} \dots}\ =             \ \int_0^\infty
  \frac{ \alpha_1^{\rm u-1} \dd\alpha_1 }{\Gamma(\rm u)} \int_0^\infty
  \frac{ \alpha_2^{\rm v-1} \dd\alpha_2 }{\Gamma(\rm v)} \dots
  \ \,\exp{ \Big( - \alpha_1 F_1 - \alpha_2 F_2 \dots \Big) }.
\,\ \end{equation}
The factors $F_1,F_2,\dots$ must have positive Real parts;
hence, before  using formula (\ref{eq:param}), we apply the $n^*_\mu$-
prescription to any factors of $n\tdot q$ and $n\tdot k$ in the
denominator, perform Wick rotations, set $\theta$ to zero, and
take the factor(s) $\nsdot n$ from \Eq(\ref{eq:ML}) outside of the
integral.  The component $p_4$ of the external momentum is regarded as
Real until after integration has been completed.

   \subsection{The Nested Method}

After parametrization, it is necessary to integrate over both the loop
momenta and the parameters.  The question of the most suitable order
for these integrations naturally arises.  To take advantage of known
one-loop results, we might try the {\em nested method}, in which we
integrate first over {\em one} of the loop momenta ($k$, say), then
over the parameters associated with factors involving $k$, then over
the other momentum $q$, and finally over the remaining parameters.

As an example, consider the divergent Euclidean-space integral
\begin{equation} \label{eq:eg1}
  I_1\,=\,\int\int \frac{ n\dotg \, \dd^{2\omega}q \, \dd^{2\omega}k }
                     { n\tdot q [(p-q)^2+m^2] [(p-q-k)^2+m^2] k^2 }\,,
\end{equation}
arising from the overlapping and rainbow diagrams.  Applying formula
(\ref{eq:param}) to the $k$-dependent factors only, we obtain
\[
  I_1\,=\,\int \frac{ n\dotg \, \dd^{2\omega}q }{ n\tdot q\,Q }
         \int_0^\infty \td\alpha_1 \int_0^\infty \td\alpha_2
        \int \td^{2\omega}k \,
       \exp{ \Big( - \alpha_1 [(p-q-k)^2+m^2] - \alpha_2 k^2 \Big) },
\]
where $\,\ Q \equiv (p-q)^2+m^2.\ \,$  Next we change variables from
$\ \alpha_1,\alpha_2\ $ to $\ A \equiv \alpha_1+\alpha_2\,$, $\,
x \equiv \alpha_1/A,\,$ and complete the square in the exponent to get
\begin{equation}
  I_1\,=\,\int \frac{ n\dotg \, \dd^{2\omega}q }{ n\tdot q\,Q }
          \int_0^1 \td x \int_0^\infty \! A\,\dd A \int \td^{2\omega}k
          \,\exp{ \Big( -A [ (k-xp+xq)^2+H ] \Big) }, \label{eq:xak}
\end{equation}
with $\,\ H \equiv x\Big[ (1-x)(p-q)^2+m^2 \Big].\ \,$  The $k$ and $A$
integrations may then be carried out with the help of the well-known
Gaussian and Gamma integrals:
\begin{equation}                                 \label{eq:Gauss}
   \int\exp{(-Ar^2)}\,\dd^{2\omega}r = \left(\frac{\pi}A
        \right)^\omega, \qquad\quad A>0,   \end{equation}
and \begin{equation}
   \int_0^\infty A^d e^{-AH} \,\dd A = \frac{\Gamma(d+1)}{H^{d+1}},
   \qquad\quad H>0,\ \ \re d>-1,                     \label{eq:Gamma}
\end{equation}

\smallskip
\noindent  respectively, with $\ r_\mu=k_\mu-x(p-q)_\mu\ $ in this
case.  Accordingly, we obtain
\begin{equation} \label{eq:nstd}
  I_1\ =\ \pi^\omega\Gamma(\eps) \int\frac{n\dotg\,\dd^{2\omega}q}
         { n\tdot q\,Q } \int_0^1 \td x \, H^{-\eps};
  \quad\qquad \eps \equiv 2-\omega.
\end{equation}
\par From
power counting, we expect the $k$ integral in \Eq(\ref{eq:eg1})
to be well defined only if $\omega < 2$.  In fact, we required just
this condition in order to complete the integration (\ref{eq:Gamma})
which produced the divergent factor $\Gamma(\eps)$ in
\Eq(\ref{eq:nstd}).   The $q$ integral in \Eq(\ref{eq:nstd}) also
diverges as $\eps\to 0$, according to power
counting, so altogether we expect $I_1$ to have a {\em double} pole at
$\eps=0$.  This expectation will be confirmed by explicit calculation.

Before we can complete the integration, we must decide how to deal with
the $q$-dependent factor $H^{-\eps}$ in \Eq(\ref{eq:nstd}).  Since
we expect $I_1$ to have a double pole at $\eps=0$, and since we are
only interested in finding the divergent parts, we might try
integrating only terms up to order $\,\eps\,$ from the exponential
series
\begin{eqnarray} \label{eq:exp}
  H^{-\eps} \!\!&=&\!\! 1 - \eps\ln H + \half (\eps\ln H)^2 -\dots ~,\\
     &=&\!\! 1-\eps\ln x - \eps\ln \Big[ (1-x)(p-q)^2+m^2 \Big]
           + \half (\eps\ln x)^2 + \dots ~,                  \\
     &=&\!\! x^{-\eps} + \Big[ (1-x)(p-q)^2+m^2 \Big]^{-\eps}
             - 1 + {\rm O}(\eps^2).   \label{eq:xex}
\end{eqnarray}
If we could drop the O$(\eps^2)$ terms,  we could immediately integrate
over $x$, and then complete the $q$ integration from
\Eq(\ref{eq:nstd}).  Unfortunately, the series (\ref{eq:exp}) cannot be
integrated term-by-term, since it does not converge {\em uniformly}
with respect to $q$, as explained in paper (I).

The convergence of series (\ref{eq:exp}) is non-uniform partly
because $H$ goes to infinity as $|q|$ goes to infinity.  One way of
solving this problem is to factor out the large $|q|$ behaviour
{\it before} using a series, provided we  do so without either
creating new convergence problems, or generating terms that we cannot
 integrate.  In the current example, we can extract a factor of $Q$
from
$H$ before using the exponential series, so that \Eqs(\ref{eq:exp}) to
(\ref{eq:xex}) become
\begin{eqnarray} \label{eq:expq}
  H^{-\eps} \!\!&=&\!\! Q^{-\eps} \left( 1 - \eps\ln\frac HQ + \frac12
            \left[ \eps\ln\frac HQ \right]^2 - \dots \right), \\
   &=&\!\! Q^{-\eps} \left( 1 - \eps\ln x - \eps\ln \left[
           \frac{ (1-x)(p-q)^2+m^2 }Q \right] + \frac12 (\eps\ln x)^2
           + \dots \right), \nonumber \\
   &=&\!\! Q^{-\eps} \left( x^{-\eps} + \left[
           \frac{ (1-x)(p-q)^2+m^2 }Q \right]^{-\eps} - 1
           + {\rm O}(\eps^2) \right),  \label{eq:xexq}
\end{eqnarray}
with $\,Q \equiv (p-q)^2+m^2$.
Convergence of this new series remains uniform as $|q|\to\infty$,
{\em except} near $x=0$ and $x=1$.  Fortunately, however, these
remaining non-uniformities cause no trouble, provided we integrate
\Eq(\ref{eq:xexq}) only.  At $x=0$, the $x^{-\eps}$ term in this
equation gives the correct contribution to the pole parts of $I_1$,
while the contributions from the second and third terms in parentheses
cancel.  Similarly, as $x\to 1$, $|q|\to\infty$, the second term in
parentheses gives the correct contribution,  while the contributions
from the first and third terms cancel.  (One may check these claims by
using expansions in powers of $x$ near $x=0$, and $1-x$ near $x=1$.)
Hence, the O$(\eps^2)$ term in \Eq(\ref{eq:xexq}) does not contribute
to the pole parts of $I_1$.

Integrating \Eq(\ref{eq:xexq}) over $x$, and substituting into
\Eq(\ref{eq:nstd}), we find that
\begin{equation}  \label{eq:nexp}
  I_1\ =\ \frac{ \pi^\omega\Gamma(\eps) }{ 1-\eps } \int\frac{
  n\dotg\,\dd^{2\omega}q }{ n\tdot q\,Q } \left[ \frac{1+\eps}{Q^\eps}
  + \frac{ m^2 \big[ Q^{-\eps}-(m^2)^{-\eps} \big] }{ (p-q)^2 }
  \right] \quad + {\rm\ finite},
\ \end{equation}
where ``finite'' refers to terms which do not diverge as $\eps\to 0$.
By power counting, we see that the $q$ integration in
\Eq(\ref{eq:nexp}) diverges only for the first term in square brackets.
 Because of the divergent factor $\Gamma(\eps)$ in front, we need
both the divergent {\em and} finite parts of the $q$ integral, but {\em
not} parts of order $\eps$.  Hence, we may set $\eps=0$ in the
convergent second term, causing this term to vanish (fortuitously).
Applying the $n^*_\mu$-prescription (\ref{eq:ML}), along with the
parametrization formula (\ref{eq:param}), we obtain
\[
  I_1 \ =\ \frac{ 2n\dotg \pi^\omega \Gamma(\eps) (1+\eps) }
     { \nsdot n \,(1-\eps) } \int_0^\infty \td\alpha_1 \int_0^\infty
     \frac{ \alpha_2^\eps\,\dd\alpha_2 }{\Gamma(1+\eps)} \int\!
     \nsdot q\,e^{-T}\,\dd^{2\omega}q\ \ + {\rm\ finite},
     \qquad\qquad \]
\begin{equation}
  =\ \frac{ 2n\dotg \pi^\omega (1+\eps) }{ \nsdot n\,\eps (1-\eps)}
     \int_0^1\! y^\eps\dd y \int_0^\infty\!\! A^{1+\eps}\dd A
     \int\! \nsdot (r+yp_\|+p_\perp)\,e^{-T}\,\dd^{2\omega}r
     \ \ + {\rm\ finite},  \label{eq:iyar}
\ \ \end{equation}

\smallskip
\noindent
where $\quad A\equiv \alpha_1+\alpha_2,
\quad y\equiv \alpha_2/A,\quad r_\mu\equiv
q_\mu-yp_{\|\mu}-p_{\perp\mu},\quad$ and
\begin{eqnarray*}
       T \,\equiv\, \alpha_1 q_\|^2 +\alpha_2 Q
    &=& A\Big[ q_\|^2+yq_\perp^2-2yp\tdot q+y(p^2+m^2) \Big], \\
    &=& A\Big[ r_\|^2+yr_\perp^2+y(1-y)p_\|^2+ym^2 \Big].
\end{eqnarray*}
The 4-vectors $p_{\|\mu},\,p_{\perp\mu}$, and so on, are defined in
\Eqs(\ref{eq:pp}).  Note that $\,\nsdot p_\|=\nsdot p$,
$\,p\tdot q=p_\|\tdot q_\|+p_\perp\tdot q_\perp,
\ (p_\|)_\perp=0,\,$ etc., for any 4-vectors $p_\mu$ and $q_\mu$.

Since $T$ is an even function of $r$, we may drop the term with
$\nsdot r$ from the integrand of \Eq(\ref{eq:iyar}).  We also drop
$\nsdot p_\perp$, which is equal to zero, and factor the $r$
integral into independent $r_\|$ and $r_\perp$ parts:
\begin{eqnarray*}
  \int y\,\nsdot p_\| \,e^{-T}\,\dd^{2\omega}r &=& y\,
  \nsdot p\,\exp{ \Big( -Ay \Big[ (1-y)p_\|^2+m^2 \Big] \Big)
  } \\ && \times\ \int \exp{( -A r_\|^2 )}\,\dd^2 r_\|
           \int \exp{( -A yr_\perp^2 )}\,\dd^{2\omega-2} r_\perp.
\end{eqnarray*}
The integrations over $r,A$, and $y$ are then easily completed with the
help of  formulas (\ref{eq:Gauss}) and
(\ref{eq:Gamma}).

   \subsection{The Quark-Bubble Integral $I_{\rm e}$}

Let us try the nested method on the integral in \Eq(\ref{eq:f}).
Since the inner loop of Figure 1e involves no gluons, the integral
over $k$ of the $k$-dependent factors in (\ref{eq:f}) is just the
standard covariant-gauge result \cite{ry}:
\begin{eqnarray}
\lefteqn{   \int_M \td^{2\omega}k \, \frac{ {\rm Trace}
   [ \gamma^\sigma (k\dotg+m_L) \gamma^\rho (k\dotg-q\dotg+m_L) ] }{
   [ k^2-m_L^2+i\theta ][ (k-q)^2-m_L^2+i\theta ] } \ = } \nonumber \\
   && \qquad\quad 8i\pi^\omega \Gamma(\eps)(q^2 g^{\sigma\rho}
      - q^\sigma q^\rho) \int_0^1 \! (x-x^2)
      [m_L^2-(x-x^2)q^2]^{-\eps}\dd x, \qquad \label{eq:vpol}
\end{eqnarray}
with $\theta>0$ and $\eps\equiv 2-\omega$ as
before.  The above equation is {\em exact}:  no terms of order
$\,\eps\,$, or higher, have been omitted.  (To verify this claim, one
may use the trace theorems \cite{ry} to obtain \newline
Trace $\![ \gamma^\sigma (k\dotg+m_L) \gamma^\rho (k\dotg-q\dotg+m_L) ]
\ =\ 4g^{\sigma\rho}(m_L^2-k^2-q\tdot k) + 8k^\sigma k^\rho - 4q^\sigma
k^\rho - 4k^\sigma q^\rho$,
and then carry out the $k$ integration using some of the formulas and
methods from Subsection 3.1.)

Substituting the right-hand sides of \Eqs(\ref{eq:vpol}) and
(\ref{eq:quark}) into \Eq(\ref{eq:f}), we get
\begin{equation} \label{eq:fxm}
   I_{\rm e} \,=\, 8i\pi^\omega \Gamma(\eps) \int_M \td^{2\omega}q
      \frac{ \gamma^\mu (\pslash-q\dotg+m) \gamma^\nu }
           { (p-q)^2-m^2+i\theta } \,D_{\mu\nu}(q) \int_0^1 \!
      \frac{ (x-x^2)\,\dd x }{ [m_L^2-(x-x^2)q^2]^\eps }\,,
\ \ \end{equation}
where $\ D_{\mu\nu}(q) \equiv G_{\mu\sigma}(q)(q^2 g^{\sigma\rho}-
q^\sigma q^\rho) G_{\rho\nu}(q),\ $ and $m$ is the rest mass of the
{\em external} quark in Figure 1e.  It follows from \Eq(\ref{eq:prop})
 that $\,D_{\mu\nu}(q) \to -G_{\mu\nu}(q)\,$ as
$\,\theta,\lambda\to 0$, so  that $I_{\rm e}$ is proportional to
the one-loop integral (\ref{eq:a}), except for the extra $x$ integral
within the $q$ integral.  We note that there is  only
one factor of $q^2$ now in the denominator (in $D_{\mu\nu}(q)$),
thereby preventing the integral from diverging as $q\to 0$.

After Wick rotation, \Eq(\ref{eq:fxm}) becomes
\begin{equation} \label{eq:fxe}
  I_{\rm e}=\,8\pi^\omega \Gamma(\eps) \int_E \td^{2\omega}q
            \,\frac{ \gamma^\mu (\pslash-q\dotg-m) \gamma^\nu }
                   { [(p-q)^2+m^2]\,q^2 }   \! \left( \delta_{\mu\nu}
            - \frac{ n_\mu q_\nu + q_\mu n_\nu }{ n\tdot q }  \right)
        \! \int_0^1 \frac{ (x-x^2)^{1-\eps} \!\! }{ L^\eps } \,\dd x,
\ \ \end{equation}
with $\,\ L \equiv q^2+m_L^2(x-x^2)^{-1}.\ \,$  As in the example of
\Eq(\ref{eq:eg1}), power counting tells us that integral
(\ref{eq:fxe}) is well defined only if $\eps>0$.  Under this condition,
$(q^2)^{-\eps}$ is a decreasing, concave-up function of $q^2$ (for
$q^2>0$), from which it follows that
\begin{equation}
   0\ <\ (q^2)^{-\eps} - L^{-\eps}\ <\ (q^2-L)
   \frac{ \dd }{ \dd (q^2) } \Big[ (q^2)^{-\eps} \Big] \quad
   =\, \frac{ \eps\,m_L^2 }{ x-x^2 } (q^2)^{-\eps-1}.  \label{eq:in}
\ \end{equation}
Using this inequality, we can show by power counting that the
difference between integral (\ref{eq:fxe}), and the same integral with
$L$ replaced by $q^2$, remains finite as $\eps\to 0$.  Since we are
merely interested in  the divergent parts of $I_{\rm e}$, we may
make this replacement and pull the factor $(q^2)^{-\eps}$ out of the
$x$ integral.  Integrating over $x$ with the help of the formula
\begin{equation} \int_0^1 x^c (1-x)^d \dd x \ = \ \frac{ \Gamma(c+1)
  \Gamma(d+1) }{ \Gamma(c+d+2) }\,,\qquad\re c,\re d>-1,\label{eq:beta}
\end{equation}
and completing the $q$ integration, we get the desired result.  Notice
that the divergent parts of $I_{\rm e}$ will be independent of the mass
of the quark in the inner loop, as
$m_L$ enters \Eq(\ref{eq:fxe}) only by way of $L$.

   \subsection{Review of the Matrix Method}

As seen in the preceding examples, in the nested method one tries to
modify the integrand between the first and second momentum
integrations, in a way that  simplifies the final integration without
changing the divergent parts of the result.  Particular care must be
taken with regard to the behaviour of the integrand near boundaries
where the final integration diverges, such as $|q|\to\infty,\,x\to 0$
in the examples.  The greater the degree of divergence, the more
closely the ``simplified'' integrand must match the exact one.  This
requirement becomes more and more challenging as the number of $\alpha$
parameters increases.

For integrals such as \begin{equation}   \label{eq:eg2} I_2 \,=\,
  \int_E\int_E\,\frac{ q\dotg\ \dd^{2\omega}q\,\dd^{2\omega}k}
   { n\tdot q\,[(p-q)^2+m^2]\,[(p-q-k)^2+m^2]\,[(p-k)^2+m^2]\,k^2 }\,,
\ \end{equation}
in which both $q$ and $k$ appear in more than two denominator factors
each, it is easier to complete {\em all} momentum integrations before
doing {\em any} parameter integrations.  This approach is called the
{\em matrix method} \cite{me,me2}.  Momentum integrations are
straightforward with this method, because the combined
$4\omega$-dimensional momentum integral can be expressed as a
derivative of a product of one-dimensional Gaussian integrals.  In view
of the importance of the matrix method for multi-loop integrals, we
shall briefly review its main features.

After  Wick rotation and application of the $n^*_\mu$-prescription, a
two-loop light-cone integral takes the form of an integral over $q$ and
$k$ of a polynomial $P(q,k)$, say, divided by some non-negative
quadratic factors $F_1,F_2,\dots$.  Application of the parametrization
formula (\ref{eq:param}) then yields
\begin{equation}                                      \label{eq:8}
   \int_E \td^{2\omega}q \int_E \td^{2\omega}k\,
   \frac{ P(q,k) }{ F_1 F_2\dots }\ =\ \int_0^\infty\td\alpha_1
   \int_0^\infty\td\alpha_2\ \dots\ J[P(q,k)];
\ \end{equation} \begin{equation}                     \label{eq:9a}
     J[P(q,k)]\ \equiv\ \int_E \td^{2\omega}q \int_E \td^{2\omega}k\,
       P(q,k)\,\exp{ \Big( -\alpha_1 F_1-\alpha_2 F_2\dots \Big) }.
\ \end{equation}

\smallskip \noindent  Since the exponent is quadratic in $q$ and $k$,
we can rewrite \Eq(\ref{eq:9a}) in the form
\begin{equation}                                      \label{eq:9b}
   J[P(q,k)]\ =\ \int_E \td^{4\omega}{\mathsf z}\,
     P(q,k)\,\exp{\mathsf \Big( -zMz^\top+2B\tdot z-C \Big) },
\ \ \end{equation}
where $\mathsf M$ is a Real $4\omega \!\times\! 4\omega$ matrix,
$\mathsf B$ and $\mathsf z$ are Real $4\omega$-vectors, ${\mathsf z}
\equiv (k_4,q_4,\,k_3,q_3$, $\dots),{\ }^\top$ denotes transpose, and
the components of $\mathsf M,\,B$, and $\mathsf C$
are functions of $\alpha_1,\alpha_2\dots$, but {\em not} of $q$ or $k$.

For any given set of $F$ factors, one may obtain explicit expressions
for the components of $\mathsf M,\,B$, and $\mathsf C$ by expressing
the exponents from \Eqs(\ref{eq:9a}) and (\ref{eq:9b}) in terms
of the components of $q$ and $k$, and then equating corresponding
coefficients.  For the two-loop integrals of the Appendix, with
$q_\|=(0,0,q_3,q_4)$ and $k_\|=(0,0,k_3,k_4)$ as before, we find that
\begin{equation}
  {\mathsf M} = A \left[ \begin{array}{ccccccc} \!
    a & tG &&&&& \\  \! tG & \,1-a\, &&&&& \\ && \!
    a & tG &&& \\ && \! tG & \,1-a\, &&& \\ &&&& \!
    h & tG & \\ &&&& \! tG \,& \lambda & \\ &&&&&& \ddots \!
     \end{array} \right], \qquad {\mathsf B}^\top = A \left[\!
   \begin{array}{c} bp_4 \\ \beta p_4 \\ bp_3 \\ \beta p_3 \\ bp_2
     \\ \beta p_2 \\ \vdots  \end{array} \!\right],   \label{eq:11}
\quad\end{equation} where
the dots in $\mathsf M$ denote $2\omega-3$ repetitions of the third
$2\!\times\! 2$ sub-matrix, and $A(1-a),A\lambda,A\beta,AG,Ab,Ah$, and
$Aa$ are the sums of the $\alpha$ parameters whose corresponding $F$
factors include $\quad q_\|^2,\quad q_\perp^2,\quad -2p\tdot q,\quad
2tq\tdot k,\quad -2p\tdot k,\quad k_\perp^2,\quad k_\|^2,\ $
respectively.  For example, if we label the denominator factors in
integral (\ref{eq:eg2}) as $F_1$ to $F_5$ from left to right (after
application of prescription (\ref{eq:ML}), with
$2\nsdot q/\nsdot n$ taken into $P$),   we have for this integral
\begin{equation} \qquad
  \ \left. \begin{array}{l}  1-a=(\alpha_1+\alpha_2+\alpha_3)/A,
          \qquad \lambda=\beta=(\alpha_2+\alpha_3)/A, \qquad  t=1, \\
  h=a=(\alpha_3+\alpha_4+\alpha_5)/A, \qquad b=(\alpha_3+\alpha_4)/A,
  \qquad G=\alpha_3/A,\end{array}\ \ \ \right\}\!\!\!\!\label{eq:ahbG}
\end{equation}
\begin{equation}   {\mathsf C}\ =\ (\alpha_2+\alpha_3+\alpha_4)
           (p^2+m^2)\ =\ A(b+\lambda-G)(p^2+m^2).  \qquad \label{eq:C}
\end{equation}
The equations for $\,a\,$ and $\,1-a\,$ tell us
that $\,A=\alpha_1+\alpha_2+2\alpha_3+\alpha_4+\alpha_5$.

\bigskip
Since the matrix $\mathsf M$ is to be multiplied by $\mathsf z$ on
both sides, it may always be constructed symmetrically.  Accordingly,
there exists a matrix $\mathsf R$ such that $\mathsf RMR^\top$ is
diagonal and $\mathsf R^\top\! R=1$.  Defining a new vector $\mathsf y$
so that $\mathsf z=yR+BM^{-1}$, and taking $P=1$, we find that
\Eq(\ref{eq:9b}) becomes
\begin{equation}                               \label{eq:14}
   J[1]\ =\ \exp{\mathsf \Big( BM^{-1}B^\top -C \Big) } \int_E
      |\!\det\mathsf R| \,\dd^{4\omega}y\,
      \exp{ \Big( -yRMR^\top\! y^\top \Big) }.
\ \ \end{equation}
Since the integral in this equation
is just a product of $4\omega$ one-dimensional Gaussian integrals,
 we may use formula (\ref{eq:Gauss}), along with
$\ \ |\!\det\mathsf R|=1\ \ $ and $\ \ \prod_i({\mathsf RMR^\top})_{ii}
=\det(\mathsf RMR^\top)=\det M,\ \,$ to get
\begin{equation}
   J[1]\ =\ \pi^{2\omega}(\det {\mathsf M})^{-1/2}
          \exp{\mathsf \Big( BM^{-1} B^\top-C \Big) }. \label{eq:15}
\end{equation}
Notice that we never actually need to construct $\mathsf R$.  To find
$J[P]$ for general $P(q,k)$, we differentiate \Eq(\ref{eq:9b})
partially with respect to ${\mathsf B}_i$ to obtain
$\,\partial J[P]/\partial {\mathsf B}_i=2J[{\mathsf z}_i P]$,
and then apply this formula repeatedly to \Eq(\ref{eq:15}) to get
\begin{equation}
   \left. \begin{array}{rcl}
   J[{\mathsf z}_i] &=& {\mathsf T}_i J[1], \\
   J[{\mathsf z}_i{\mathsf z}_j] &=& \Big[ {\mathsf T}_i{\mathsf T}_j
       + (\half {\mathsf M}^{-1})_{ij} \Big] J[1], \\
   J[{\mathsf z}_i{\mathsf z}_j{\mathsf z}_k] &=& \Big[
       {\mathsf T}_i{\mathsf T}_j{\mathsf T}_k +
       {\mathsf T}_i(\half {\mathsf M}^{-1})_{jk} +
       {\mathsf T}_j(\half {\mathsf M}^{-1})_{ki} +
       {\mathsf T}_k(\half {\mathsf M}^{-1})_{ij} \Big] J[1], \ \\
       &\vdots& \end{array}   \right\}    \label{eq:18}
\end{equation} \begin{equation}
    {\rm where}\quad {\mathsf T}\ \equiv\ {\mathsf BM}^{-1}\ \equiv
    \ (r_4,s_4,\ r_3,s_3,\,\dots\,). \qquad\qquad\qquad \label{eq:19}
\end{equation}
Since the momentum integral $J[P]$ is a linear functional of $P$, and
$\,{\mathsf z}\equiv (k_4,q_4,\,k_3,q_3$, $\dots)$, we can use the
above equations to find $J[P]$ for any polynomial $P(q,k)$.

\bigskip
In order to derive the two-loop integrals summarized in the Appendix,
we substitute from \Eqs(\ref{eq:11}) into \Eqs(\ref{eq:15}) to
(\ref{eq:19}) to obtain the following relations:
\begin{eqnarray}
  && \quad J[k_\mu]\,=\,r_\mu J[1],\quad\qquad J[\nsdot q\,k_\mu]\,=
     \,\left( \nsdot s\,r_\mu-\frac{Gn^*_\mu}{2AD_\|} \right) J[1],
     \qquad \nonumber \\
  && \quad J[q_\mu]\,=\,s_\mu J[1],\quad\qquad J[\nsdot q\,q_\mu]\,=
     \,\left( \nsdot s\,s_\mu+\frac{an^*_\mu}{2AD_\|} \right) J[1],
     \qquad \label{eq:J} \\
  && J[\nsdot q\,n\tdot k\,k_\mu]\ =\ \left(\nsdot s\,n\tdot r\,r_\mu
     +\frac{ \beta\nsdot p n_\mu - G\nsdot n(r+r_\|)_\mu }{ 2AD_\| }
     \right) J[1], \qquad \nonumber
\end{eqnarray}
\begin{equation} {\rm and\ so\ on,\ where } \qquad     \label{eq:J1}
   J[1] \,=\, \left( \frac{\pi}A \right)^{2\omega} \frac{ e^{-AH} }
   { D_\| D_\perp^{\omega-1} }\,, \qquad\qquad\qquad\qquad\qquad\qquad
\end{equation} \begin{eqnarray}
   r_\mu \!&=& [(1-a)b-G\beta] \frac{p_{\|\mu}}{D_\|}  \label{eq:r}
      + (\lambda b - G\beta) \frac{ p_{\perp\mu} }{ D_\perp }\,, \\
   s_\mu \!&=& \qquad (a\beta-Gb) \frac{p_{\|\mu}}{D_\|}
      + (h\beta-Gb) \frac{ p_{\perp\mu} }{ D_\perp }\,,\label{eq:s} \\
   H    \! & \equiv & \frac{ \mathsf C-BM^{-1}B^\top }A
      \ =\,\frac{ \mathsf C }A - (br+\beta s)\tdot p, \label{eq:HH} \\
   D_\| \! & \equiv & a(1-a)-G^2, \qquad\quad
      D_\perp \,\equiv\ \lambda h-G^2.                 \label{eq:D}
\end{eqnarray} From the
definitions of the new parameters following  \Eq(\ref{eq:11}), one
may show that the sub-determinants $D_\|$ and $D_\perp$ satisfy $D_\|
\geq 0$, $D_\perp \geq 0$ for all allowed values of $a,\lambda,h$, and
$G$, thereby justifying our use of formula (\ref{eq:Gauss}) in the
derivation of \Eq(\ref{eq:15}).

For the integral (\ref{eq:eg2}), the polynomial $P [q,k]$ is given by
$P=2\nsdot q q\dotg/\nsdot n$. Since $h=a$ and $\beta=\lambda$ from
\Eqs(\ref{eq:ahbG}), it follows from \Eqs(\ref{eq:J}), (\ref{eq:J1}),
(\ref{eq:s}), (\ref{eq:HH}), and (\ref{eq:C}) that
\begin{eqnarray}  J[P] \!&=&\!
   \frac2{ \nsdot n } \left[ \nsdot p \frac{ (a\lambda-Gb)^2 }{ D_\| }
   \left( \frac{ \pslash_\| }{ D_\| } + \frac{ \pslash_{\!\perp} }
   { D_\perp }\right) + \frac{ a\,n\dotg^* }{ 2AD_\| } \right] \frac{
   \pi^{2\omega} e^{-AH} }{ A^{2\omega} D_\| D_\perp^{\omega-1} }\,;
      \ \label{eq:JP} \\         \label{eq:H}
   H \!&=&\! (b+\lambda-G)(p^2+m^2) - (br+\lambda s)\tdot p.
\end{eqnarray}

   \subsection{Parameter Integration for the Matrix Method }

Once $J[P]$ is known for a particular integral, we must complete the
parameter integrations in \Eq(\ref{eq:8}).  For the two-loop integrals
listed in the Appendix, we begin these integrations by changing
variables from $\alpha_1,\alpha_2,\dots$ to the applicable subset of
the new parameters $\lambda,\beta,G,b,h,a$ and $A$, defined after
\Eq(\ref{eq:11}).  For the sample integral (\ref{eq:eg2}), we find
from \Eqs(\ref{eq:ahbG}) that
\begin{equation}
   \int_0^\infty\td\alpha_1 \int_0^\infty\td\alpha_2\,\dots\,J\ \,=
   \ \int_0^{1/2}\td G \int_G^{1-G}\td a \int_G^{1-a}\td\lambda
   \int_G^a\td b \int_0^\infty \! JA^4\dd A.       \label{eq:fp}
\ \end{equation}
The integration ranges of the ``finite'' parameters
$\lambda,\beta,G,b,h$, and $a$ depend on which $F$ factors are present
in the original integral.  For instance, if the factor containing $q+k$
in \Eq(\ref{eq:eg2}) did not also contain $p$,  the $b$ integration in
\Eq(\ref{eq:fp}) would run from $0$ to $a-G$, rather than from $G$ to
$a$. In the event of a {\it repeated} $F$ factor, there will be an
additional finite parameter which can be  integrated
out immediately, since $J$ will not depend on it.

We see from \Eqs(\ref{eq:D}) and (\ref{eq:H}) that
$D_\|,\,D_\perp$, and $H$ are independent of $A$, so that the $A$
integration in \Eq(\ref{eq:fp}) is straightforward.  Applying
 formula (\ref{eq:Gamma}) to \Eq(\ref{eq:JP}), we obtain
\begin{equation}
  \int_0^\infty\! J[P] A^4\dd A\ =\ \frac{ \pi^{2\omega} }{ \nsdot n }
   \Big[ \Gamma(1+2\eps)J_1 + \Gamma(2\eps)J_0 \Big];\ \quad
   \eps \equiv 2-\omega,                                 \label{eq:JA}
\end{equation} where
\begin{equation}                  \label{eq:J01}
  J_1 \,=\,\frac{ 2\nsdot p\, (a\lambda-Gb)^2 }{ D_\|^2
     D_\perp^{1-\eps} H^{1+2\eps} } \left( \frac{ \pslash_\| }{D_\|}
     + \frac{ \pslash_{\!\perp} }{ D_\perp } \right), \qquad
  J_0 \,=\,\frac{ n\dotg^*a H^{-2\eps} }{ D_\|^2 D_\perp^{1-\eps} }\,.
\ \end{equation}

\smallskip \noindent  Note that the
use of formula (\ref{eq:Gamma}) requires $H>0$.  We can see
that this requirement is satisfied in general by examining the way in
which $H$ was constructed: since the $F$ factors and $\alpha$
parameters in  \Eq(\ref{eq:8}) are non-negative, we find that the
exponents in (\ref{eq:9a}) and (\ref{eq:9b}) are $\leq 0$ for {\it all}
values of $q$ and $k$.  Consequently, the {\it sum} of the exponents in
\Eq(\ref{eq:14}) is $\leq 0$ for all values of $\mathsf y$.  Taking
$\mathsf y=0$, and noting that $A$ is just a sum of $\alpha$
parameters, we see that $H$  in \Eq(\ref{eq:HH}) is $\geq 0$ for all
allowed values of the old or new parameters.  Furthermore, since
$r_\mu$ and $s_\mu$ are independent of the quark mass $m$, one can
deduce from \Eq(\ref{eq:H}) that  $H=0$ only if $\,b+\lambda-G=0$
($m\neq 0$).  We emphasize that the relationship $\,b+\lambda-G=0\,$
applies specifically to the integral (\ref{eq:eg2}).

\bigskip  The presence of the divergent factor $\Gamma(2\eps)$ in
\Eq(\ref{eq:JA}) reflects the UV-divergence of integral (\ref{eq:eg2})
with respect to $q$ and $k$ combined.  Additional divergences, known as
{\em subdivergences}, may be found when we complete the remaining
integrations from \Eq(\ref{eq:fp}), because $J_0$ and $J_1$ go to
infinity at the boundaries of the integration region where $a\to 0$ and
$a\to 1$ (corresponding to $|k|\to\infty$ and $|q|\to\infty$,
respectively).  In the current example, each of these boundaries has
three fewer dimensions than the full finite-parameter space, because
from the limits of integration in (\ref{eq:fp}) it follows that
\begin{equation}
   a\ \geq\ b\ \geq\ G\ \geq\ 0\qquad {\rm and} \qquad 1-a\ \geq
   \ \lambda\ \geq\ G\ \geq\ 0.                       \label{eq:lim}
\end{equation}
Thus, near $a=1$, for instance, we could transform $dG d\lambda$
to $(1-a)^2 dx dy$ (with $x$ and $y$ being {\em finite} parameters
such that $\lambda=x(1-a)$ and $G=y(1-a)$).  A subdivergence can
therefore occur at $a=1$ only if the integrand $J_0$ or $J_1$ diverges
there at least as fast as $(1-a)^{-3}$.
\par From
\Eqs(\ref{eq:D}), (\ref{eq:J01}), and (\ref{eq:lim}), we
find that $D_\|,D_\perp\to 0$ linearly as $a\to 1$, while the numerator
of $J_1$ goes to zero quadratically, and $H$ remains positive unless
$b\to 0$ as well (assuming $m\neq 0$, as discussed above).  Thus, we
find that $J_0$ is of order $(1-a)^{\eps-3}$ near $a=1$, $b\neq 0$,
while $J_1$ is of order $(1-a)^{\eps-2}$ there.  Hence, by the
criterion given above, the integral of $J_0$ over the finite parameters
diverges as $a\to 1$ (and $\eps\to 0$), while the integral of $J_1$
does not.  Similar analyses at other boundaries of the integration
region (including the case $a=1$, $b=0$) show that there are no other
subdivergences in this example.  Since we are only interested in
finding the divergent parts of $I_2$, and since $J_1$ gives no
subdivergences and is not multiplied by a divergent Gamma function in
\Eq(\ref{eq:JA}), we may drop $J_1$.

\bigskip
Due to the factor $\Gamma(2\eps)$ in \Eq(\ref{eq:JA}), the
subdivergence in the integral of $J_0$ will contribute a double pole to
$I_2$, while the {\it finite part} of the same integral will contribute
to the single pole.  The factor $H^{-2\eps}$ in \Eq(\ref{eq:J01})
complicates the integration of $J_0$, but some simplification is
possible because this factor affects the pole parts of $I_2$ only by
way of the subdivergence at $a=1$.  To begin the simplification, we
rewrite $H^{-2\eps}$ in (\ref{eq:J01}) as $1+(H^{-2\eps}-1)$, and then
expand the factors, multiplying the term $(H^{-2\eps}-1)$, in powers
of $1-a$, $\lambda$, and $G$, using the definitions  for $D_\|$
and $D_\perp$ from \Eq(\ref{eq:D}).  In this fashion we obtain
\begin{equation}
  J_0\ =\ \frac{ n\dotg^*a }{ D_\|^2 D_\perp^{1-\eps} }
        + \frac{ n\dotg^* (H^{-2\eps}-1) }{ (1-a)^2 \lambda^{1-\eps} }
        \Big[ 1+{\rm O}(1-a,\lambda,G) \Big].   \label{eq:J0a}
\end{equation}
By counting powers of $1-a,\lambda$, and $G$, we see that the integral
of the term involving $(H^{-2\eps}-1)$O$(1-a,\lambda,G)$ has no
subdivergence;  to find its {\it finite part} we therefore can set
$\eps=0$ in this term (which happens to make it vanish).  Accordingly,
we may replace $J_0$ by $\widehat J_0(H)$,
\begin{equation}                            \label{eq:J0h}
  \widehat J_0(H)\ \equiv\ \frac{ n\dotg^*a}{ D_\|^2 D_\perp^{1-\eps}}
    + \frac{ n\dotg^* (H^{-2\eps}-1) }{ (1-a)^2 \lambda^{1-\eps} }\,,
\end{equation}
without affecting the divergent parts of $I_2$.

The next step is to replace $H$ in \Eq(\ref{eq:J0h}) by
\[   H_1\ \equiv\ \lim_{a\to 1} H\ =\ b(p^2+m^2)-b^2p^2,
\]
where the result on the right follows from \Eqs(\ref{eq:H}),
(\ref{eq:r}), (\ref{eq:D}), (\ref{eq:lim}), and from the equality
$\beta=\lambda$ in \Eqs(\ref{eq:ahbG}). To justify this replacement, we
use the same kind of reasoning that led to the inequality
(\ref{eq:in}); in the present case, an appropriate inequality is given
by \[
    0\ \ \leq\ \ |H^{-2\eps}-H_1^{-2\eps}|\ \ \leq\ \ 2\eps |H-H_1|
    \,(H^{-2\eps-1}+H_1^{-2\eps-1}),   \]
which holds for all allowed values of the parameters.  Exploiting this
new inequality, and noting that $\vert H-H_1\vert$ goes to zero as
$a\to 1$, and that both $H$ and $H_1$ go to zero (linearly) only when
$b\to 0$, we can show that the integral of $\,\widehat J_0(H)-
\widehat J_0(H_1)\,$, over the finite parameters, has no subdivergence
and vanishes when $\eps\to 0$.  In summary, we can replace $H$ by $H_1$
in \Eq(\ref{eq:J0h}) {\it without affecting the divergent parts of
$I_2$}.

Making this replacement, and integrating over $b$ in accordance with
\Eq(\ref{eq:fp}), we obtain
\begin{equation}                                \label{eq:Jb}
   \int_G^a \widehat J_0(H_1)\dd b\ =\ \frac{ n\dotg^*a(a-G) }
      { D_\|^2 D_\perp^{1-\eps} } + \frac{ n\dotg^*\lambda^{\eps-1} }
      { (1-a)^2 } \! \left[ \int_0^1\!-\!\int_0^G\!-\!\int_a^1 \right]
      \!(H_1^{-2\eps}-1)\dd b.
\ \end{equation}
As $a\to 1$, the ranges of the second and third integrals in square
brackets shrink to points, so  that these terms contribute no
subdivergences to the total integral over the {\em remaining}
parameters.  Hence, in the derivation of the {\it finite parts} of the
integrals of these terms, we are allowed to set $\eps=0$ in their
integrand (causing it to vanish). To facilitate the  remaining $b$
integration in \Eq(\ref{eq:Jb}), we shall express $H_1^{-2\eps}$ in the
form (cf.\,\Eq(\ref{eq:xex})),
 \[
  H_1^{-2\eps}\ =\ [b(p^2+m^2)-b^2p^2]^{-2\eps}
  \ =\ (p^2+m^2-bp^2)^{-2\eps}+b^{-2\eps}-1+{\rm O}(\eps^2). \]
It now remains to integrate the factor multiplying the $b$ integral
in \Eq(\ref{eq:Jb})  over $\lambda,a$, and $G$, in
accordance with \Eq(\ref{eq:fp}). But this task is easy: since
the $b$ integral is of order $\eps$,  only the {\it divergent} part of
the integral over $\lambda,a$, and $G$ is needed.

\bigskip
Finally, we must also integrate the first term on the right-hand side
of \Eq(\ref{eq:Jb})  over $\lambda,a$, and $G$, again in accordance
with \Eq(\ref{eq:fp}).  Because of the factor $\Gamma(2\eps)$ in
\Eq(\ref{eq:JA}), both the divergent and finite parts of this triple
integral will be needed.  We begin the integration process by defining
the new variables $\ x=D_\perp/a^2,\ y=D_\|/a^2,\ $ and $\ v=G/a,\,$
so that \[
    \int_0^{1/2}\td G \int_G^{1-G}\td a \int_G^{1-a}\td\lambda
    \frac{ a(a-G) }{ D_\|^2 D_\perp^{1-\eps} }\ =\ \int_0^1\td v
    \int_{v-v^2}^\infty \td y \int_{v-v^2}^y
    \frac{ a^{2\eps}(1-v) }{ y^2\,x^{1-\eps} }\dd x.
    \] From
the definitions of $y,\,v$, and $D_\|$, we find that
$\,a^{2\eps}=(y+v^2+1)^{-2\eps}$.  But this factor can be simplified
because it affects the divergent and finite parts of the integral only
at subdivergences.  The subdivergence  at $a=1$  now manifests itself
at $\,v=y=x=0$.  Near this point, we have
\begin{equation}  \label{eq:s1}
  (y+v^2+1)^{-2\eps}\ =\ 1-2\eps(y+v^2)+\eps(1+2\eps)(y+v^2)^2-\dots\,.
\end{equation}
Only the first term of this series gives a subdivergence, so we may set
$\eps=0$ in the other terms (causing them to vanish, as usual).  Away
from the subdivergence, the integral is finite, so we can take $\eps=0$
in that region as well.  Thus, in the current example, the factor
$a^{2\eps}$ may be replaced by 1.  The integrations over $x,\,y$, and
$v$ pose no further problems.

For a subdivergence at $a=0$, we would use \[
   (y+v^2+1)^{-2\eps}\ =\ y^{-2\eps} \Big[ 1 - 2\eps(v^2+1)/y +
   \eps(1+2\eps)(v^2+1)^2/y^2 - \dots \Big], \]  valid for large $y$.
If the subdivergence comes only from the first term in this series,
then $a^{2\eps}$ may be replaced by $y^{-2\eps}$.

   \subsection{Reduction of Subdivergences }

The subdivergence in integral (\ref{eq:eg2}) is of the mildest possible
nature; that is, $J_0$ goes to infinity only just fast enough to make
the finite-parameter integral diverge as $a\to 1$ and $\eps\to 0$.
For this reason, we were able to drop the O$(1-a)$ term in
\Eq(\ref{eq:J0a}), replace $H$ by $H_1$, drop the last two integrals in
\Eq(\ref{eq:Jb}), and drop all but the first term of series
(\ref{eq:s1}).  For an integral with a more severe subdivergence, it
may be helpful to begin with a partial calculation by the nested
method, thereby reducing the degree of divergence with respect to one
of the loop momenta before applying the full matrix method.

A good example is the  Euclidean-space integral
\begin{equation}
   I_3 \,=\,\int\int\frac{ P_{\mu\nu}(q,n,p)\ k^\mu k^\nu
      \,\ \dd^{2\omega}q\,\dd^{2\omega}k }{ [(q-k)^2+m^2]\,[k^2+m^2]\,
                       F_3(q) F_4(q) \dots }\,, \label{eq:eg3}
\end{equation} in which
$P_{\mu\nu}$ and $F_3,F_4,\dots$ do {\em not} depend on $k$.  Because
of the {\it quadratic} subdivergence as $|k|\to\infty$, a direct
application of the matrix method  would lead to a finite-parameter
integral proportional to $\int a^{\eps-2}\dd a$ near $a=0$ (after all
parameters except $a$ had been integrated out).  To avoid having to
deal with such a severe subdivergence, we proceed in the spirit of
\Eq(\ref{eq:xak}) by parametrizing the
$k$-dependent denominator factors only, to get
\begin{equation}                                    \label{eq:I3}
  I_3 \,=\,\int \frac{ P_{\mu\nu}\dd^{2\omega}q }{ F_3 F_4\dots }
           \int_0^1 \td v \int_0^\infty \! AY[k^\mu k^\nu]\dd A,
\end{equation} where
\begin{equation}                                    \label{eq:Y}
   Y[f(k)] \,\equiv\, \int\td^{2\omega}k\,f(k)
               \exp \Big( -A[k^2-2vq\tdot k+vq^2+m^2] \Big).
\end{equation}

\smallskip \noindent
Since \Eq(\ref{eq:Y}) has the same form as \Eq(\ref{eq:9b}) (with
$\ {\mathsf z}=k,\ {\mathsf M}=A,\ {\mathsf B}=Avq,\ $ and
$\ {\mathsf C}=A(vq^2+m^2)$), application of \Eqs(\ref{eq:15}) and
(\ref{eq:18}) yields
\[   Y[1] \,=\, \left( \frac\pi{A} \right)^\omega e^{-AL}, \qquad
     Y[k^\mu k^\nu] \,=\,\Big[ v^2 q^\mu q^\nu + \half\delta^{\mu\nu}
              A^{-1} \Big] \left( \frac\pi{A} \right)^\omega e^{-AL},
\]
with $\,\ L\equiv (v-v^2)q^2+m^2.\ \,$
We next use formula (\ref{eq:Gamma}) and \Eq(\ref{eq:Y}) to obtain
\begin{equation}
    \int_0^\infty\! AY[k^\mu k^\nu]\dd A
    \ =\ \frac{ \pi^\omega\Gamma(\eps)E^{\mu\nu} }{ L^\eps }
    \,=\, E^{\mu\nu} \int_0^\infty\! AY[1]\dd A \\
    \,=\, \int_0^\infty\! AY[E^{\mu\nu}]\dd A,    \label{eq:AYE}
\ \end{equation} where
$\ E^{\mu\nu} \equiv v^2 q^\mu q^\nu + L\delta^{\mu\nu}/(2\eps-2).\ \,$

{} From the left and right ends of \Eq(\ref{eq:AYE}), we see that $I_3$
is unchanged if we replace the
$k$-dependent argument $k^\mu k^\nu$ in \Eq(\ref{eq:I3})
 by the parameter-dependent polynomial $E^{\mu\nu}(q,v)$,
thereby reducing the degree of the subdivergence.  We then cancel $F$
factors, wherever possible, with factors in the terms of $P_{\mu\nu}
E^{\mu\nu}$, parametrize the remaining $F$ factors in
accordance with formula (\ref{eq:param}), and finally complete the
integration by using either the matrix method or the nested method,
whichever is easier to apply.

%
\bigskip
\section{Renormalization}

Wherever a quark line appears in a physical process, any one of the
diagrams of Figure 1, as well as higher-loop diagrams, could also
appear.  Hence, the {\em effective} quark propagator
$\,i\delta_{\alpha\beta}S_{\rm eff}$ is the sum of the bare propagator
$\,i\delta_{\alpha\beta}S\,$ from Subsection 2.1, plus contributions
from self-energy processes with one loop, two loops, and so on:
\begin{eqnarray}                                         \label{eq:iS}
   iS_{\rm eff}  \!&=&\!  iS + iS \Sigma_1 iS + iS \Big(
   \Sigma_1 iS \Sigma_1 + \Sigma_2 \Big) iS + \dots, \\  &=&\! iS + iS
   \Big( \Sigma_1 + \Sigma_2 + \dots \Big) iS_{\rm eff}, \label{eq:re}
\end{eqnarray}
where $\Sigma_1$ is the one-loop amputated Green function defined by
\Eq(\ref{eq:a}), $\Sigma_2$ is the sum of the Green functions for the
two-loop diagrams of Figures 1a to 1e, and so on.  ($\Sigma_2$ will
also include amplitudes of one-loop processes with counterterm
vertices, as explained below.)  Since $\Sigma_j$ is proportional to
$g^{2j}$, $j=1,2,3,\dots$, the right-hand side of (\ref{eq:iS})
is a power series in $g^2$.  Similar series may be constructed for the
effective gluon propagator and vertex factors:
\begin{eqnarray}
  i\delta^{ab} (G_{\rm eff})_{\mu\nu} \!&=&\! i\delta^{ab} G_{\mu\nu} +
   iG_{\mu\sigma} \Big[ \Pi_{(1)}^{ab\sigma\rho} +
   \Pi_{(2)}^{ab\sigma\rho} \Big] iG_{\rho\nu} +\dots, \label{eq:gr} \\
  (\Gamma_{\rm eff})^{a\mu}_{\alpha\beta} \!&=&\!
   ig\gamma^\mu T^a_{\alpha\beta} +\Big[ \Gamma^{a\mu}_{(1)\alpha\beta}
   + \Gamma^{a\mu}_{(2)\alpha\beta} \Big] + \dots,   \label{eq:vr}
\end{eqnarray} etc.,
where $\Pi_{(1)}$, $\Pi_{(2)}$, $\Gamma_{(1)}$, and $\Gamma_{(2)}$ are
the amputated Green functions of the one-loop processes shown in
Figures 2b to 2e.  For the light-cone gauge, we have, in Minkowski
space \cite{L3,B1,ry},
\begin{eqnarray}
  \Sigma_1(p) \!&=&\! -i \widetilde\alpha_s C_F         \label{eq:sig1}
      \left( 2m - \pslash + \frac{ n\dotg n\dotg^* \pslash +
    \pslash n\dotg^* n\dotg }{ \nsdot n } \right)\ \ + {\rm finite}, \\
  \Gamma^{a\mu}_{(1)\alpha\beta} \!&=&\! i\widetilde\alpha_s g
    T^a_{\alpha\beta} \left( \frac{N_C}2 - C_F \right)  \label{eq:gam1}
    \left( \gamma^\mu + 2\frac{ n\dotg\,n^{*\mu} - n\dotg^*\,n^\mu }
            { \nsdot n } \right)\ + {\rm finite},  \\   \label{eq:gam2}
  \Gamma^{a\mu}_{(2)\alpha\beta}(k) \!&=&\! -i\widetilde\alpha_s g
    T^a_{\alpha\beta} \frac{N_C}2  \left( \gamma^\mu - 2\frac{ n\dotg\,
    n^{*\mu} + n\dotg^* n^\mu }{ \nsdot n } + 4\frac{ \nsdot k\,n\dotg
    \,n^\mu }{ n\tdot k \,\nsdot n } \right)\ + {\rm finite},   \\
  \Pi_{(2)}^{ab\mu\nu}(q) \!&=&\! i\widetilde\alpha_s \delta^{ab} N_C
    \left[ \frac{11}3 \Big( q^2g^{\mu\nu}-q^\mu q^\nu \Big)
    + \frac{ 2n\tdot q }{ \nsdot n }               \nonumber
    \left( \left[ q^\mu - \frac{ q^2 n^\mu }{ n\tdot q } \right]
        \left[ n^{*\nu} - \frac{ \nsdot q\,n^\nu }{ n\tdot q } \right]
    \right.\right. \\     && \qquad\qquad\qquad + \,\left.\left.
       \left[ q^\nu - \frac{ q^2 n^\nu }{ n\tdot q } \right]
    \left[ n^{*\mu} - \frac{ \nsdot q\,n^\mu }{ n\tdot q } \right]
    \right) \right]\ + {\rm finite},  \label{eq:pi2} \\  \label{eq:pi1}
  \Pi_{(1)}^{ab\mu\nu}(q) \!&=&\! -i\widetilde\alpha_s \delta^{ab}
    \frac23 N_f \Big( q^2g^{\mu\nu}-q^\mu q^\nu \Big)\ +{\rm finite};
\end{eqnarray}
here, $p,\,q$, and $k$ are external
momenta as shown in Figure 2, $C_F=$ ``color factor'' $=\frac43\ \,
(\delta_{\alpha\kappa}C_F\equiv T^a_{\alpha\beta}T^a_{\beta\kappa}),
\ N_f=$ number of quark flavors $=6,\ N_C=$ number of colors $=3\ \,
(\delta^{ad} N_C \equiv f^{abc}f^{bcd}),\,$ and                     \[
      \widetilde\alpha_s \equiv g^2\Gamma(2-\omega)(4\pi)^{-\omega} \]
in the ``modified minimal subtraction'' scheme, $\overline{\rm MS}$.

\begin{figure}
\epsfxsize = 12cm
\epsffile{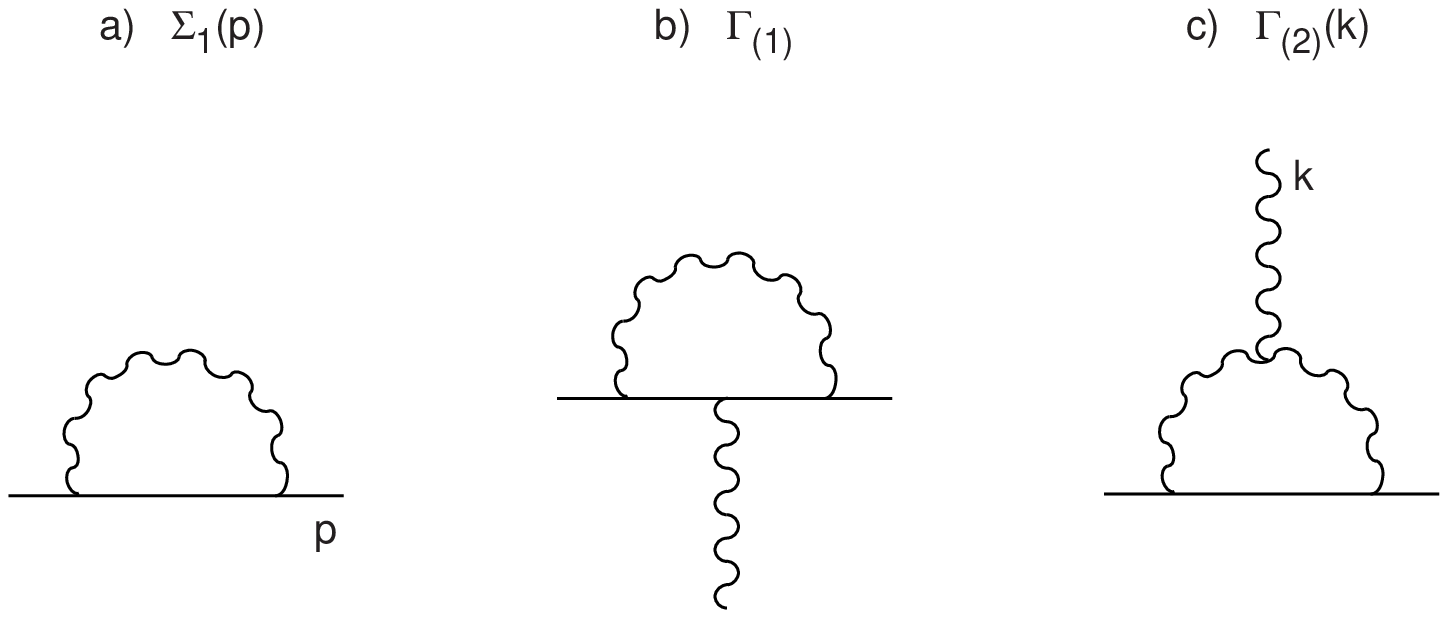}
%
\begin{picture}(200,110)(0,0)
\epsfxsize = 12cm
\epsffile{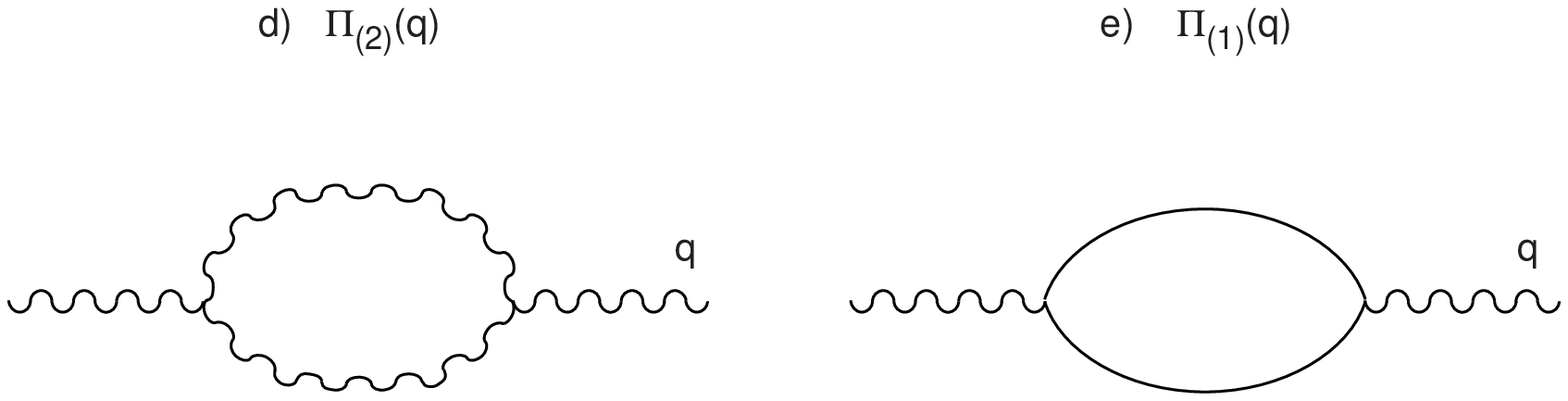}
\end{picture}
\caption{
One-loop subgraphs of Figures 1a to 1e, with their amplitudes.
 Wavy lines denote gluons, straight lines quarks. $p,\,q$, and $k$ are
four-momenta.}
\end{figure}
\clearpage
Most of the terms in \Eqs(\ref{eq:iS}), (\ref{eq:gr}), and
(\ref{eq:vr}) diverge as $\omega\to 2$.  To make the effective
propagators and vertex factors {\em finite}, we  modify the Lagrangian
density (\ref{eq:L}) in such a way that the {\em bare}
propagators and vertex factors $S$, $G$, $ig\gamma^\mu$, etc.~will be
replaced by {\em renormalized} propagators and vertex factors.
For instance, if we let
\begin{equation}  m\,\ \to\,\ m-\delta m \quad\qquad
     {\rm and} \quad\qquad \psi\ \,\to\,\ \psi+N\psi  \label{eq:mpr}
\end{equation}
in \Eq(\ref{eq:L}), where $N$ and $\delta m$ are both of order $g^2$,
then the quark propagator in \Eq(\ref{eq:quark}) becomes
\begin{equation}                                         \label{eq:sr1}
   S_{\rm ren}(p) \,=\,\Big[ (1 + \gamma^0 N^\dagger \gamma^0)
              (\pslash-m+\delta m) (1+N) \Big]^{-1},   \end{equation}
\begin{equation}
   =\, S(p) - S(p) \Big[ \delta m + \gamma^0 N^\dagger \gamma^0
   (\pslash-m)+(\pslash-m)N \Big] S(p)\ +{\rm O}(g^4), \label{eq:sr2}
\ \end{equation}

\smallskip \noindent where the $i\theta$ term has been suppressed  for
clarity. At the same time, we see from \Eq(\ref{eq:iS}) that in order
to make $S_{\rm eff}$ finite, we need
to replace $S$ by
\begin{equation}        S_{\rm ren} \,=\, S - iS \Big[ {\rm divergent
   \ part\ of}\ \Sigma_1 \Big] S\ +{\rm O}(g^4)+\dots\!. \label{eq:sr3}
\end{equation}
Comparing \Eqs(\ref{eq:sr2}) and (\ref{eq:sr3}), substituting for
$\Sigma_1$ from \Eq(\ref{eq:sig1}), and noting that $\,n\dotg^\dagger =
n\dotg^*\,$ and $\,\gamma^0 n\dotg^* = n\dotg \gamma^0$, we find that
we can eliminate the one-loop divergence from the effective quark
propagator by taking \cite{B1}
\begin{equation}                           \label{eq:dmN}
   \delta m \,=\, 3m\widetilde\alpha_s C_F + {\rm O}(g^4)  \qquad
   {\rm and} \qquad  N \,=\, \widetilde\alpha_s C_F  \left( \frac{
   n\dotg^* n\dotg }{ \nsdot n } - \frac12 \right) + {\rm O}(g^4).
\quad\end{equation}
Similarly, to eliminate {\em two-loop} divergences from $S_{\rm eff}$,
we will need to compute $\Sigma_2$,
and so on.

The substitutions $\,m\to m-\delta m\,$ and $\,\psi\to\psi+N\psi\,$
produce the necessary {\em counterterms} in the Lagrangian density
(\ref{eq:L}).  From \Eq(\ref{eq:dmN}) we see that the counterterms
involving $N$ are noncovariant, but we may hide this noncovariance by
absorbing the factor $(1+N)$ into the {\it normalization} of the quark
field $\psi$.  Similar renormalizations of the gluon fields $A^a_\mu$
lead to cancellation of the noncovariant divergences in the effective
gluon propagator and three- and four-gluon vertex corrections.
Bassetto, Dalbosco, and Soldati (BDS) have pointed out some time ago
that noncovariant divergences may be eliminated in this way from the
effective light-cone propagators and vertex factors, at {\em all
orders} of perturbation theory \cite{B2,B1}.  We shall verify this
claim explicitly for the case of the effective quark propagator at
two-loop order.

   \subsection{One-loop Counterterm Subtractions }

The use of renormalized propagators and vertex factors in place of $S$,
$G$, $ig\gamma^\mu$, etc.~affects $S_{\rm eff}$ through the $\Sigma$
factors in \Eq(\ref{eq:iS}), as well as through $S$ directly, because
the $\Sigma$ factors involve amplitudes which depend on the propagators
and vertex factors through the Feynman rules.  In accordance with
standard procedure, these ``indirect" contributions to $S_{\rm eff}$
are represented by {\it  subtraction diagrams}, such as those
depicted in Figure~3.
The diagrams of Figure~3 are specifically constructed from those of
Figs.\,1a to 1e  by collapsing the divergent one-loop subgraphs
(shown in Fig.\,2) to the counterterm vertices denoted by $\times$ in
Figure~3.  In each case, the counterterm vertex factor  is {\it minus}
the divergent part of the amplitude of the corresponding subgraph.
These amplitudes are given in \Eqs(\ref{eq:sig1}) to (\ref{eq:pi1}).
The terms proportional to $n^\mu$ or $n^\nu$ in \Eqs(\ref{eq:gam2}) and
(\ref{eq:pi2}) may be dropped (see Figs.\,3c and 3d), since they are
orthogonal to the light-cone propagators of the gluons.

\begin{figure}[h]
\begin{center}
\epsfxsize = 14cm
\epsffile{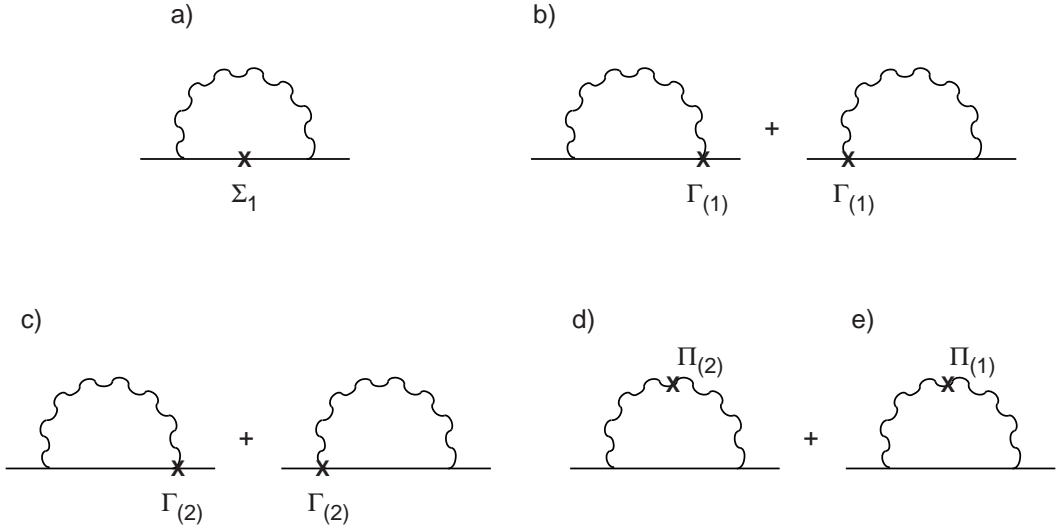}
\caption{\ One-loop subtraction diagrams.  The vertex factors for
   ``$\times$'' are {\em minus} the divergent parts of the indicated
          one-loop amplitudes $\Sigma_1,\,\Gamma_{(1)}$, etc.}
\end{center}
\end{figure}

\begin{figure}[h!]
\begin{center}
\epsfxsize = 14cm
\epsffile{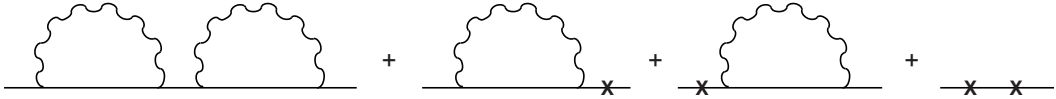}
\caption{One-particle-reducible quark self-energy diagrams.
The vertex factor for $\times$ is given by [$-$ divergent
part of $\Sigma_1$.]}
\end{center}
\end{figure}

The one-loop amplitudes $\Sigma_1,\,\Gamma_{(1)},\,\Gamma_{(2)},\,
\Pi_{(2)}$, and $\Pi_{(1)}$ are  of order $g^2$, so that the
amplitudes of Figures 3a to 3e are all of order $g^4$.  These
amplitudes belong, therefore, to $\Sigma_2$, along with the amplitudes
of Figures 1a to 1e, of course.  In principle, the amplitudes of
the one-particle-{\em reducible} diagrams (Figure 4) also
contribute to $\Sigma_2$, but it is easy to see that the divergent
parts of these amplitudes cancel completely with one another, due
to the factorizability of all four integrals.

All of the diagrams of Figures 3 and
1a to 1e contribute nonlocal divergences to $\Sigma_2$; however, the
nonlocal contributions from Figs.\,1a to 1e are exactly cancelled by
the nonlocal parts from the corresponding diagrams of Figure~3, as
expected on theoretical grounds \cite{co,sm,CB}.

\subsection{Two-loop Counterterms }

The divergent parts of the amplitudes for Figures 1a to 1e and 3a to 3e
are shown in Table 1, excluding the nonlocal divergent terms which
cancel as noted above.  To find the local divergent part of the
amplitude for a particular figure, one multiplies the numbers in the
applicable row of the table by the corresponding factors at the top of
the table, then adds the resulting terms together, and multiplies the
sum by the color factor at the right-hand end of the row.  (We recall
that $\,\eps\equiv 2-\omega$, and $\,\widetilde\alpha_s \equiv
g^2\Gamma(\eps)(4\pi)^{-\omega}$ for modified minimal subtraction.)
The divergent part of $\Sigma_2$ is the sum of the results for all ten
figures, as shown in the ``total'' section at the bottom of the table.
\clearpage

\begin{table}[h]  \caption[ Local divergent parts of the amplitudes for
  Figures 1a to 1e and 3a to 3e. ]{Local divergent parts of the
  amplitudes for Figures 1a to 1e and 3a to 3e.  Each number is to be
  multiplied by the expressions at the top of its column and right-hand
  end of its row. $C_F,\,N_C,\,N_f$, and
  $\,\widetilde\alpha_s\,$ are defined after \Eq(67).  Columns with 1
  at the top give double-pole terms; columns with $\eps$, single poles.
  Vectors are in Minkowski space, and $\eps \equiv 2-\omega$. }

\begin{tabular} { |l @{\ \ \ }r @{\,}|r r|r r|r r|r r|l @{} }
  \multicolumn{2}{c}{\ } & \multicolumn{2}{c}{
   $\stackrel{\displaystyle m}{\overbrace{\rule{ 42pt}{0pt}}} $ } &
   \multicolumn{2}{c}{ $\stackrel{\displaystyle \pslash}{\overbrace{
                                                \rule{ 42pt}{0pt}}} $ }
 & \multicolumn{2}{c}{ $\stackrel{\displaystyle \frac
  { \nsdot p\,n\dotg }{ \nsdot n } }{\overbrace{\rule{ 42pt}{0pt}}} $ }
 & \multicolumn{2}{c}{ $\stackrel{\displaystyle \frac
  { n\tdot p\,n\dotg^* }{\nsdot n} }{\overbrace{\rule{ 42pt}{0pt}}} $ }
 & \\ [-8pt]        \multicolumn{2}{|c|}{ Figure\ \ }
 & 1 & $\eps$ & 1 & $\eps$ & 1 & $\eps$ & 1 & $\eps$ & \\ \hline\hline
 rainbow & 1a & --8 &  16 & --3 & --$\frac32$ & --6 &--40 &   6 &   8 &
    $\frac12 i\widetilde\alpha_s^2 C_F^2$ \\
         & 3a &  16 &--24 &   6 &           2 &  12 &  24 &--12 & --8 &
    $\frac12 i\widetilde\alpha_s^2 C_F^2$ \\  \hline
 over-   & 1b &   4 &   8 &   4 &           2 &  14 &  60 & --6 &--12 &
    $\frac12 i\widetilde\alpha_s^2 (C_F^2-\frac12 N_C C_F) $ \\ [-8pt]
 lapping &&&&& &&&&& \\ [-12pt]
         & 3b & --8 &   0 & --8 &         --4 &--28 &--40 &  12 &   8 &
    $\frac12 i\widetilde\alpha_s^2 (C_F^2-\frac12 N_C C_F) $ \\  \hline
 spider  & 1c &   4 &--16 &   0 &         --2 & --6 &--20 & --2 &   4 &
    $\frac14 i\widetilde\alpha_s^2 N_C C_F $ \\
         & 3c & --8 &   0 &   0 &         --4 &  12 &  24 &   4 &   8 &
    $\frac14 i\widetilde\alpha_s^2 N_C C_F $ \\  \hline
 gluon-  & 1d  &    --$\frac{44}3$  &--$\frac{124}9$  &--$\frac{10}3$ &
   --$\frac{47}9 $  & $\frac{16}3$  &  $\frac{284}9$  &  $\frac{32}3$ &
     $\frac{292}9$ & $ \frac14 i\widetilde\alpha_s^2 N_C C_F $\\ [-8pt]
 bubble  &&&&& &&&&& \\ [-12pt]
         & 3d  &  $\frac{88}3$ &  0 &  $\frac{20}3 $  &  $\frac{44}3$ &
   --$\frac{32}3 $ &--$\frac{104}3$ &--$\frac{64}3 $  &--$\frac{88}3$ &
     $ \frac14  i\widetilde\alpha_s^2  N_C C_F $  \\  \hline
 quark-  & 1e  &      $\frac83   $  &  $\frac{40}9 $  &  $\frac43   $ &
     $\frac{14}9 $  & $\frac83   $  &  $\frac{64}9 $  &--$\frac83   $ &
   --$\frac{64}9 $ & $ \frac14 i\widetilde\alpha_s^2 N_fC_F $ \\ [-8pt]
 bubble  &&&&& &&&&& \\ [-12pt]
         & 3e  &--$\frac{16}3$ &  0 &--$\frac83    $  &--$\frac83   $ &
   --$\frac{16}3 $ &--$\frac{16}3$  &  $\frac{16}3 $  &  $\frac{16}3$ &
     $ \frac14  i\widetilde\alpha_s^2  N_f C_F $  \\  \hline\hline
         &     &    --$\frac83   $  &  $\frac{40}9 $  &--$\frac43   $ &
   --$\frac{10}9 $ &--$\frac83   $  &  $\frac{16}9 $  &  $\frac83   $ &
   --$\frac{16}9 $ & $ \frac14 i\widetilde\alpha_s^2 N_fC_F $\\ [-16pt]
 \ \ total $\ \left\{ \rule{0pt}{32pt} \right.\!\!\!\!\!\!\! $
           &         & $\frac{44}3$ &--$\frac{340}9$  &  $\frac{22}3$ &
     $\frac{49}9 $ &  $\frac{44}3$  &--$\frac{172}9$  &--$\frac{44}3$ &
     $\frac{172}9$ & $\frac14 i\widetilde\alpha_s^2 N_C C_F $\\ [-16pt]
 \ \ \ $(\Sigma_2)$ && 4 & 0 & --1 & --$\frac32$  & --8 & 4 & 0 & --4 &
     $\frac12 i\widetilde\alpha_s^2 C_F^2$ \\  \hline\hline
\end{tabular} \end{table}

A knowledge of the divergent part of $\Sigma_2$ enables us  to verify
explicitly that the effective quark propagator may indeed be rendered
finite at two-loop order by means of the substitutions
$\,m\to m-\delta m\,$ and $\,\psi\to \psi+N\psi\,$, as implied by
Bassetto, Dalbosco and Soldati. The O$(\widetilde\alpha_s)$ parts of
$N$ and $\delta m$ are already shown in \Eqs(\ref{eq:dmN}). Multiplying
\Eq(\ref{eq:re}) by $S^{-1}$ from the left and by
$(S_{\rm eff})^{-1}$ from the right, then solving for
$(S_{\rm eff})^{-1}$ and replacing $S$ by $S_{\rm ren}$, we get
\begin{equation}
   (S_{\rm eff})^{-1} \,=\, (S_{\rm ren})^{-1} - i\Sigma_1 - i\Sigma_2
   - \dots\!.                                           \label{eq:sei}
\end{equation}
To keep $S_{\rm eff}$ finite at two-loop order, we must ensure that the
divergent parts of $\Sigma_2$ in \Eq({\ref{eq:sei}) are cancelled by
terms from $(S_{\rm ren})^{-1}$; the latter depends on $N$ and
$\delta m$, as seen from \Eq(\ref{eq:sr1}).  (The divergent parts of
$\Sigma_1$ cancel already, thanks to the particular choice of the
O$(\widetilde\alpha_s)$ parts of $N$ and $\delta m$.)  Inverting and
expanding the right-hand side of (\ref{eq:sr1}), we obtain
\begin{equation}                                        \label{eq:sri}
   (S_{\rm ren})^{-1} \,=\, S^{-1} + \delta m + (\delta m - m)
    \Big( N^*+N+N^*N \Big) + N^*\pslash N + N^*\pslash + \pslash N,
\ \ \end{equation}
where $\,N^*\equiv\gamma^0 N^\dagger\gamma^0 $.  Comparison of the
terms of this expansion with the expressions at the tops of the columns
in Table 1 suggests that the O$(\widetilde\alpha_s^2)$ parts of $N$ and
$\delta m$ should be similar in form to the O$(\widetilde\alpha_s)$
parts shown in \Eqs(\ref{eq:dmN}).  Therefore, let us try the ansatz
\begin{equation}
   \delta m \,=\, 3m\widetilde\alpha_s C_F + m\widetilde\alpha_s^2
       C_F W\ + {\rm O}(\widetilde\alpha_s^3),     \label{eq:bob}
\end{equation} and
\begin{equation}
   N \,=\, \widetilde\alpha_s C_F \left( \frac{ n\dotg^* n\dotg }
      { \nsdot n } - \frac12 \right) + \widetilde\alpha_s^2 C_F
      \left( X \frac{ n\dotg^* n\dotg }{ \nsdot n } + Y \right)\ +
      {\rm O}(\widetilde\alpha_s^3),                 \label{eq:joe}
\end{equation}
and then see if we can find expressions for $W,\,X$, and $Y$ that will
cause all O$(\widetilde\alpha_s^2)$ terms in \Eq(\ref{eq:sei}) to
cancel.

Substituting from \Eqs(\ref{eq:bob}) and (\ref{eq:joe}) into
\Eq(\ref{eq:sri}), and applying the light-cone condition
$\,n^2=n^{*2}=0$, we find that the noncovariant
O$(\widetilde\alpha_s^2)$ part of $(S_{\rm ren})^{-1}$ reads
\begin{equation}
   \widetilde\alpha_s^2 C_F \left[ (2X-C_F) \frac{\nsdot p \,n\dotg
    - n\tdot p\,n\dotg^* }{ \nsdot n } + 4 C_F \frac{\nsdot p \,n\dotg}
    { \nsdot n } \right].         \label{eq:fred}
\end{equation}
$S_{\rm eff}$ can be finite only if there is some value of $X$ that
makes expression (\ref{eq:fred}) match the noncovariant, divergent part
of $i\Sigma_2$.  Such a match is possible, but only because the
coefficients under $\nsdot p\,n\dotg$ in the ``total'' part of Table 1
are the negatives of the corresponding coefficients under
$n\tdot p\,n\dotg^*$, {\em except} for the last two double-pole
coefficients in the last row.  These last double-pole coefficients give
rise to a term which cancels with the last term in brackets in
(\ref{eq:fred}).  This term in (\ref{eq:fred}) derives purely
from the {\em leading} term of $N$, via the $N^*\pslash N$ term in
\Eq(\ref{eq:sri}), so its coefficient cannot be adjusted (unlike $X$)
to make it cancel with some arbitrary term from $i\Sigma_2$.
Consequently, the anti-symmetry between the coefficients of
$\nsdot p\,n\dotg$ and $n\tdot p\,n\dotg^*$ in $\Sigma_2$ must be
broken in exactly the way that we have found it to be broken, otherwise
the BDS scheme would have failed.

To complete our derivation of $N$ and $\delta m$, having substituted
from \Eqs(\ref{eq:bob}) and (\ref{eq:joe}) into \Eq(\ref{eq:sri}), we
then substitute from \Eq(\ref{eq:sri}) and Table 1 into
\Eq(\ref{eq:sei}), and set the total of the O$(\widetilde\alpha_s^2)$
terms on the right to zero.  In this way we find the following
expressions for $W$, $X$, and $Y$:
\begin{eqnarray}                  \label{eq:defW}
 W \!&=&\! \frac{ N_f }4 \left(       4 - \frac{10}3  \eps \right)
         + \frac{ N_C }4 \left(     -22 + \frac{97}3  \eps \right)
         + \frac{ C_F }2 \left(       -9 + \frac32    \eps \right), \\
 X \!&=&\! \frac{ N_f }4 \left(   \frac43 - \frac89   \eps \right)
         + \frac{ N_C }4 \left( -\frac{22}3 + \frac{86}9 \eps \right)
         + \frac{ C_F }2 \left( 1 - 2 \eps \right),  \label{eq:defX} \\
 Y \!&=&\! \frac{ N_f }4 \left( -\frac23  + \frac{13}9   \eps \right)
         + \frac{ N_C }4 \left( \frac{11}3-\frac{221}{18}\eps \right)
         + \frac{ C_F }2 \left( \frac14   + \frac{11}4   \eps \right).
\label{eq:defY} \end{eqnarray}
Thus we have confirmed the claim of Bassetto, Dalbosco, and Soldati for
the two-loop quark self-energy function.
It is also significant to realize that the mass counterterm $\delta m$
in the light-cone gauge, \Eq(\ref{eq:bob}), is exactly the same as in
the general linear covariant gauge
\cite{TF}, at least up to two-loop order.  This result is a reassuring
reflection of the gauge symmetry of QCD \cite{BX}.

%
\bigskip
\section{Conclusion}

In this paper, we have evaluated the complete quark propagator to
two-loop order, and demonstrated for the first time its
renormalizability in the noncovariant light-cone gauge. The chief
technical tool in this three-year project was the powerful {\it matrix
integration technique} which had been exploited in our first paper
\cite{me} to evaluate the divergent part of the overlapping
self-energy function (see Fig.\,1b in the present article).  Now, three
years later, we have finally succeeded in computing, without
approximation, the divergent parts of the four remaining two-loop
diagrams in the quark propagator, namely: the quark-bubble diagram
(Fig.\,1e), the rainbow diagram (Fig.\,1a), the spider diagram
(Fig.\,1c), and the gluon-bubble diagram (Fig.\,1d), along with their
counterterm graphs, depicted in Figs.\,3e, 3a, 3c and 3d, respectively.

Other technical tools, apart from the matrix method and a reliable
computer algebra program, included dimensional regularization, the
$n^*_\mu$-prescription for the spurious poles of $(q\tdot n)^{-1}$
($n^2 = 0$), as well as a detailed analysis of the boundary
singularities in five- and even six-dimensional parameter space.

The divergent part of the total two-loop correction $\Sigma_2$ to the
quark propagator can be read off from Table 1, and has the form
\begin{eqnarray}
  \Sigma_2 \ &=&\ i \widetilde\alpha^2_s \left\{ C_F^2 \left( 2m -
     \frac{ \pslash }2 - 4\,{\nsdot p\,n\dotg \over \nsdot n} \right)
     \right. \nonumber \\
  &&\ \ \left.\qquad +~\left( \frac{11}3 N_C -\frac23 N_f \right) C_F
     \left( m + \frac12 \pslash + {\nsdot p\, n\dotg - n \tdot p
     \, n\dotg^* \over \nsdot n} \right) \right\}         \nonumber \\
  && +~i \widetilde\alpha_s^2 \,\eps \left\{ C_F^2 \left(
     - \frac34 \pslash + 2 \left[ {\nsdot p\,n\dotg - n\tdot p\,
     n\dotg^* \over \nsdot n} \right] \right) \right.     \nonumber \\
  && \qquad\qquad +~\frac{ N_C C_F }9 \left( -85m + \frac{49}4
     \pslash - 43 \left[ {\nsdot p \,n\dotg - n\tdot p\,n\dotg^*
     \over \nsdot n} \right] \right)                      \nonumber \\
  && \left.\qquad\qquad +~\frac{ N_f C_F }9 \left( 10m - \frac52
     \pslash + 4 \left[ {\nsdot p \,n\dotg - n\tdot p\,n\dotg^*
     \over \nsdot n} \right] \right) \right\},        \label{eq:final}
\end{eqnarray}
with $\widetilde\alpha_s \equiv g^2 \Gamma(\eps )(4\pi)^{\eps-2}$.
The expression (\ref{eq:final}) has several noteworthy features:
\begin{itemize}
\item[(i)] The anti-symmetry between the coefficients of $\nsdot p
   n\dotg$ and $n\tdot p n\dotg^*$ (proportional to $\widetilde
   \alpha^2_s C_F^2)$ is identical to the anti-symmetry predicted by
   the Bassetto-Dalbosco-Soldati renormalization scheme \cite{B2}.
\item[(ii)] The divergent part of $\Sigma_2$ is a {\it local} function
   of the external momentum $p$, even off mass-shell, since all {\it
   nonlocal} divergent terms cancel exactly. Notice that $\Sigma_2$
   contains both covariant and noncovariant components.
\item[(iii)] The structure of $\Sigma_2$ implies the quark mass
   counterterm $\delta m$,
   \begin{equation}
      \delta m \,=\, 3 m \widetilde\alpha_s \, C_F + m
      \widetilde\alpha^2_s \,C_F\,W + {\rm O}(\widetilde\alpha^3_s)\,,
   \end{equation}
   which is gauge-independent, as expected ($W$ is given by
   \Eq(\ref{eq:defW})).  It is both interesting, and re-assuring from
   a calculational point of view, that the mass counterterm in the {\it
   noncovariant} light-cone gauge is exactly the same as in the general
   linear {\it covariant} gauge \cite{TF} -- at least to two-loop
   order.
\item[(iv)] The factor $N$ for the renormalization of the quark
   field $\psi$ is likewise {\it local}, albeit gauge-dependent:
   \begin{equation}
      N \,=\, \widetilde\alpha_s \,C_F \left( {n\dotg^* n\dotg \over
           \nsdot n} - {1\over 2} \right) + \widetilde\alpha^2_s \,C_F
           \left( X {n\dotg^* n\dotg \over \nsdot n} - Y \right)
         + {\rm O}(\widetilde\alpha^3_s)\,,
   \end{equation}
   with $X$ and $Y$ defined in \Eqs(\ref{eq:defX}) and (\ref{eq:defY}),
   respectively.
\end{itemize}

The cancellation of nonlocal divergences, the consistency of our
results with the Bassetto-Dalbosco-Soldati renormalization scheme, and
the agreement between our mass renormalization and the two-loop {\em
covariant} result all serve to demonstrate the reliability of the
matrix integration method, as well as the validity, at two-loop order,
of the $n^*_\mu$-prescription for the unphysical poles of
the light-cone gauge propagator.

Since physically meaningful predictions can only be made with a theory
whose divergences have been subtracted consistently, our
result will help to make the light-cone gauge a more useful tool for
practical two-loop calculations in QCD and other non-Abelian theories.
Complete two-loop renormalization will also allow the computation of
certain three-loop quantities, by means of {\em renormalization group
improvements} \cite{VP}.

Of course, the cancellation of the divergences in the effective quark
propagator is only {\em part} of the overall renormalization picture.
In order to complete the two-loop renormalization of light-cone QCD,
we must also find the counterterms which eliminate the two-loop
divergences from the effective gluon propagator and vertex factors.
We should be able to find these new counterterms using the same methods
that have worked in this paper for the renormalization of the quark
propagator.

%
\newpage
\begin{center}  \section*{Appendix}  \end{center}

The tables in this Appendix show one possible way in which the
integrals $I_{\rm a}$ to $I_{\rm e}$ from Subsection 2.2 may be broken
down into simpler integrals.  Specifically, if we regard the columns of
the tables as Cartesian vectors, then $I_{\rm a}$, for example, is the
sum over all five tables of the dot product of the column headed by
$I_{\rm a}$ (if any), times the integral over $q$ and $k$ of the
``integrand'' column (plus some finite integrals which have been
discarded).  Blank spaces in the tables denote zeros.  For example,
\[
  I_{\rm c}\,=\,4\int \frac{ n\dotg n\tdot p \,\dd^{2\omega} q
              \,\dd^{2\omega} k }{ \mathsf nqQx\ o }
   +2\eps\int \frac{ n\dotg \,\dd^{2\omega} q \,\dd^{2\omega} k }
              { \mathsf n\ \ xKk } +\dots\!. \]
The symbols in
the denominators of the integrands are to be decoded as follows:
\begin{eqnarray*}
  {\mathsf n} \to n\tdot q,      \ &&  {\mathsf q} \to q^2,    \qquad
  {\mathsf Q} \to [(p-q)^2+m^2],\qquad {\mathsf x} \to [(p-q-k)^2+m^2],
  \\   {\mathsf o} \to n\tdot k, \ &&  {\mathsf k} \to k^2,    \qquad
  {\mathsf K} \to [(p-k)^2+m^2],\qquad {\mathsf y} \to (q-k)^2.
\end{eqnarray*}
Note that all 4-vectors in this Appendix are in Euclidean space.

\bigskip
For each row of each table, the divergent part of the integral
over $q$ and $k$ of the expression in the ``integrand'' column is
$\,\pi^{2\omega} \Gamma(2\eps) (m^2+p^2)^{-2\eps} \,$ times the
expression in the final column, where $\,\eps\equiv 2-\omega$, and
\begin{eqnarray*}
  && b\equiv \frac{m^2}{p^2} \ln\left( \frac{ m^2+p^2 }{m^2} \right)-1,
       \qquad\qquad\  n_p \equiv \frac{ \nsdot p }{ \nsdot n } n,
       \qquad\quad\ n_p^* \equiv \frac{ n\tdot p }{ \nsdot n } n^*, \\
  && c\equiv \left( 2+\frac{m^2}{p^2} \right) b, \ \ \qquad\qquad
     h\dotg \equiv (b-f)\left( \frac{ p^2n\dotg }{ 2n\tdot p }
            - \pslash \right) - f n\dotg_p^* - n\dotg_p + un\dotg, \\
  && f \equiv \frac{ m^2+p_\|^2 }{ p_\perp^2 } \ln \left( \frac
         { m^2+p^2 }  { m^2+p_\|^2 } \right)-1,  \qquad\qquad  t\equiv
     \frac{ f\nsdot p }{ \nsdot n } + \frac{ (b-f)p^2 }{ 2n\tdot p },\\
  && a_\mu \equiv \frac{ m^2 p_{\perp\mu} }{ p_\perp^2 }
       \ln \left( \frac{ m^2+p^2 }{ m^2+p_\|^2 } \right), \qquad\qquad
     p_\| = n_p+n_p^*, \qquad\quad  p_\perp \equiv p-p_\|, \\
  && u\equiv \frac{ (m^2+p^2)d }{ 2 n\tdot p }, \,\ \qquad\qquad\qquad
     d\equiv {\rm Li}_2 \left( \frac{p^2}{ m^2+p^2 } \right)
           - {\rm Li}_2 \left( \frac{p_\perp^2}{ m^2+p^2 } \right), \\
  && v\equiv \frac{ m^2d }{ n\tdot p },   \ \qquad\qquad
     {\rm Li}_2(z) \equiv \int_0^1 \frac{ \ln\tau\ \dd\tau }
             { \tau-1/z }\ \quad {\rm (the\ ``dilogarithm")}.
\end{eqnarray*}

\begin{table} \[
\begin{array} { |@{\,} c @{\,}|@{\,} c @{\,}|@{\,} c @{\,}| r @{/} l
                | l } I_{\rm a} & I_{\rm b} & I_{\rm c} &
\multicolumn{2}{c|}{{\rm integrand},\ R} & \displaystyle \frac{
   (m^2+p^2)^{2\eps} \!\!\!\! }{ \pi^{2\omega} \Gamma(2\eps) }\,\cdot\,
   {\rm div\!.part} \!\int\! R \,\dd^{2\omega} q \,\dd^{2\omega} k  \\
\hline\hline
-4 &&& m^2 n\dotg & \mathsf nqQx\ k   & 2vn\dotg \\
\hline
2 & -4 & 4 & n\dotg n\tdot p & \mathsf nqQx\ \ o &
  4 ( 1 + b + d ) n\dotg_p - 4(t+u) n\dotg \\
& 4 && n\dotg n\tdot p & \mathsf nqQ\ K\ o & 4d n\dotg_p \\
\hline
-2\eps & 2\eps & 2\eps & n\dotg\ \ & \mathsf n\ \ xKk &
   ( \frac1\eps + 4 - 4b + 2c ) n\dotg_p \\
2 & -2-2\eps & -2 & n\dotg\ \ & \mathsf n\ Qx\ k &
   ( \frac2\eps + 4 ) n\dotg_p - 4t n\dotg \\
& -2\eps & -2\eps & n\dotg\ \ & \mathsf n\ Q\ Kk &
   ( \frac4\eps + 4 - 4b ) n\dotg_p - 4t n\dotg \\
\hline
& -4+4\eps & -4+4\eps & n\dotg q\tdot k & \mathsf nqQ\ Kk &
   ( \frac1\eps - c ) n\dotg_p + (u-t) n\dotg  \\
\hline
4-4\eps &&& n\dotg p\tdot k & \mathsf nqQx\ k &
   ( -\frac1{2\eps} - \frac12 ) n\dotg_p + (t + u - v) n\dotg  \\
\hline
4 &&& n\dotg p\tdot q & \mathsf nq\ xKk &
   ( \frac1\eps + 3 - 2b ) n\dotg_p \\
& -4 && n\dotg p\tdot q & \mathsf n\ QxKk &
   ( \frac1\eps + 3 - 2b ) n\dotg_p \\
\hline
8-8\eps & -4+4\eps & -8+8\eps & k\dotg n\tdot k & \mathsf nq\ xKk &
   -\frac58 \pslash + \frac14 n\dotg_p - \frac34 n\dotg_p^* \\
& -4+4\eps && k\dotg n\tdot k & \mathsf n\ QxKk & -\frac38 \pslash +
   ( \frac1{2\eps}+\frac34 ) n\dotg_p - tn\dotg - \frac34 n\dotg_p^* \\
\hline
&&& \multicolumn{2}{c|}{\ } & \\ [-18pt]
& 4-4\eps && \multicolumn{2}{c}{ \displaystyle \frac{ (m^2+p^2)
   k\dotg n\tdot k }{ \mathsf nqQxKk } } & un\dotg \\ [-18pt]
&&& \multicolumn{2}{c|}{\ } & \\
\hline
-4\eps & 4\eps & 4\eps & q\dotg n\tdot k & \mathsf nq\ xKk &
   -\frac14 \pslash + ( \frac1{2\eps} +1-c) n\dotg_p^* \\
& 4-4\eps && q\dotg n\tdot k& \mathsf nqQxK\ & (-\frac1{4\eps}-\frac18
   + \frac12 c ) \pslash + 2h\dotg + ( \frac1\eps +3) n\dotg_p^* \\
4-4\eps & -8+8\eps & -8+8\eps & q\dotg n\tdot k & \mathsf nqQx\ k &
   ( -\frac1{4\eps} - \frac38 + \frac12c ) \pslash
   + h\dotg + ( \frac1{2\eps} + \frac32 ) n\dotg_p^* \\
& 4-4\eps & 4-4\eps & q\dotg n\tdot k & \mathsf nqQ\ Kk &
   h\dotg + ( \frac1\eps + 1 - c ) n\dotg_p^* \\
\hline
-16+8\eps & -4+4\eps & 24-8\eps & k\dotg n\tdot p& \mathsf nq\ xKk &
   - n\dotg_p^* \\
& 20-8\eps && k\dotg n\tdot p& \mathsf n\ QxKk & - n\dotg_p^* \\
& -4+4\eps & -4+4\eps & k\dotg n\tdot p & \mathsf nqQx\ k & d\pslash
   - h\dotg - ( \frac1{2\eps} + \frac32 ) n\dotg_p^* \\
& 4-4\eps & 4-4\eps & k\dotg n\tdot p & \mathsf nqQ\ Kk & d\pslash \\
\hline
4\eps & 4-4\eps & 4-4\eps & q\dotg n\tdot p & \mathsf nq\ xKk &
   ( \frac1\eps + 3 - 2b ) n\dotg_p^* \\
& -4 && q\dotg n\tdot p & \mathsf n\ QxKk &
   ( \frac1\eps + 3 - 2b ) n\dotg_p^* \\
& 4+8\eps && q\dotg n\tdot p & \mathsf nqQxK\ & 2h\dotg +
   ( \frac1\eps + 3 ) n\dotg_p^* \\
-8 & 4-4\eps & 4 & q\dotg n\tdot p & \mathsf nqQx\ k & 2h\dotg +
   ( \frac1\eps + 3 ) n\dotg_p^* \\
& -4+4\eps & -4+4\eps & q\dotg n\tdot p & \mathsf nqQ\ Kk &
   2h\dotg + ( \frac2\eps + 4 - 2b ) n\dotg_p^* \\
\hline\hline
\end{array}   \]  \end{table}

\begin{table} \[
\begin{array} { | c | c | c | r @{/} l | l }
              I_{\rm a} & I_{\rm b} & I_{\rm c} &
\multicolumn{2}{c|}{{\rm integrand},\ R} & \displaystyle
   \frac{ (m^2+p^2)^{2\eps} \!\!\!\! }{ \pi^{2\omega} \Gamma(2\eps) }
                \,\cdot\, {\rm div\!.part} \!\int\! R \\
\hline\hline
8\eps\pslash+ & 8(1-\eps)\pslash+ & -16\eps\pslash+ & \multicolumn{2}
{c|}{\ } \\ [-14pt] &&& n\tdot k & \mathsf nq\ xKk & -1 \\ [-14pt]
8(\eps-1)m & 8(1-\eps)m & 16(1-\eps)m & \multicolumn{2}{c|}{\ } \\
\hline
& 4m-4\pslash && n\tdot k & \mathsf n\ QxKk & -1 \\
\hline
& 8(1-\eps)\pslash+ & 8(1-\eps)\pslash+ & \multicolumn{2}{c|}{\ } \\
[-14pt] &&& n\tdot k & \mathsf nqQx\ k & -\frac1{2\eps}-\frac32+b+d \\
[-14pt] & 8(1-\eps)m & 8(1-\eps)m & \multicolumn{2}{c|}{\ } \\
\hline
& 8(\eps-1)\pslash+ & 8(\eps-1)\pslash+ & \multicolumn{2}{c|}{\ } \\
[-14pt] &&& n\tdot k & \mathsf nqQ\ Kk & d \\
[-14pt] & 8(\eps-1)m & 8(\eps-1)m & \multicolumn{2}{c|}{\ } \\
\hline\hline
& 2(2-\eps)\pslash+ & 4(\eps-1)\pslash+ & \multicolumn{2}{c|}{\ } \\
[-14pt] &&& \ 1\ & \mathsf q\ xKk & \frac1\eps + 3 - 2b \\
[-14pt] & 2(1-\eps)m & 4\eps m & \multicolumn{2}{c|}{\ } \\
\hline
& 8(1-\eps)\pslash+ && \multicolumn{2}{c|}{\ } \\
[-14pt] &&& \ 1\ & \mathsf qQxK\ & \frac1\eps + 3 - 2b \\
[-14pt] & 8m && \multicolumn{2}{c|}{\ } \\
\hline
4\eps\pslash+ & 2(2-\eps)\pslash+ & 8(1-\eps)\pslash+ & \multicolumn{2}
{c|}{\ } \\ [-14pt] &&& \ 1\ & \mathsf qQx\ k & \frac1\eps + 3 - 2b \\
[-14pt] 8(\eps-1)m & 2(1-\eps)m & 8(1-\eps)m &\multicolumn{2}{c|}{\ }\\
\hline
& 2(\eps-2)\pslash+ & -2(1+\eps)\pslash & \multicolumn{2}{c|}{\ } \\
[-14pt] &&& \ 1\ & \mathsf qQ\ Kk & \frac2\eps + 4 - 4b \\
[-14pt] & 2(\eps-1)m & -2\eps m & \multicolumn{2}{c|}{\ } \\
\hline
-4\eps & 4-4\eps & 4-4\eps & q\dotg\ & \mathsf q\ xKk &
   ( \frac1{4\eps} + \frac98 - b + \frac12c )\pslash \\
& -8+8\eps && q\dotg\ & \mathsf \ QxKk &
   ( \frac3{4\eps} + \frac{23}8 - 2b + \frac12c )\pslash \\
& -8+4\eps && q\dotg\ & \mathsf qQxK\ &
   ( \frac1{2\eps} + \frac14 - c )\pslash \\
8-4\eps & -8 & -8+4\eps & q\dotg\ & \mathsf qQx\ k &
   ( \frac1{2\eps} + \frac34 - c )\pslash \\
& 4-4\eps & 4-4\eps & q\dotg\ & \mathsf qQ\ Kk &
   ( \frac1\eps + 1 - b - c )\pslash \\
\hline
4-8\eps &&& \ q\dotg\ & \mathsf qQQx & -( 2-2b+2c )\pslash \\
-8+8\eps &&& m^2 q\dotg\ & \mathsf qQQx\ k & ( 2-2b+2c )\pslash \\
-4+8\eps &&& \ m\ & \mathsf qQQx & -( 2+2b )m  \\
8-8\eps &&& \ m^3 & \mathsf qQQx\ k & ( 2+2b )m  \\
\hline\hline
\end{array}  \]  \end{table}

\begin{table} \[
\begin{array} { | c | c | c | r @{/} l | l }
              I_{\rm a} & I_{\rm b} & I_{\rm c} &
\multicolumn{2}{c|}{{\rm integrand},\ R} & \displaystyle \frac{
   (m^2+p^2)^{2\eps} \!\!\!\! }{ \pi^{2\omega} \Gamma(2\eps) }\,\cdot\,
   {\rm div\!.part} \!\int\! R \\
\hline\hline
-4 && -8 & n\dotg\,n\tdot p\,p\tdot k & \mathsf nqQx\ ko & 2dn\dotg_p\\
\hline
& 2+4\eps && (m^2+p^2) n\dotg & \mathsf nqQxK\ & 4un\dotg \\ -2 &
-2+2\eps & -2+4\eps & (m^2+p^2) n\dotg & \mathsf nqQx\ k & 4un\dotg \\
& 4-4\eps & 4-4\eps & (m^2+p^2) n\dotg & \mathsf nqQ\ Kk & 4un\dotg \\
\hline
-4+4\eps &&& q\dotg n\tdot p & \mathsf nqQQx & -( 2+2b )\pslash
   + 2a\dotg + vn\dotg \\
8 &&& m^2 q\dotg n\tdot p & \mathsf nqQQx\ k & ( 2+2b )\pslash
   - 2a\dotg - vn\dotg \\
\hline
2-2\eps &&& n\dotg & \mathsf n\ QQx & 2\Big[ a\tdot p-(b+1)p^2 \Big]
   n\dotg / n\tdot p \\
-4 &&& m^2 n\dotg & \mathsf n\ QQx\ k & 2\Big[ (b+1)p^2-a\tdot p \Big]
   n\dotg / n\tdot p \\
-2+2\eps &&& (m^2+p^2) n\dotg & \mathsf nqQQx &
   2\Big[ a\tdot p-(b+1)p^2 \Big] n\dotg / n\tdot p \\
4 &&& m^2 (m^2+p^2) n\dotg & \mathsf nqQQx\ k &
   2\Big[ (b+1)p^2-a\tdot p \Big] n\dotg / n\tdot p \\
\hline\hline
\end{array}  \]  \end{table}

\begin{table} \[
\begin{array} { | c | c | r @{/} l | l } I_{\rm d} & I_{\rm e} &
   \multicolumn{2}{c|}{{\rm integrand},\ R} & \displaystyle \frac{
   (m^2+p^2)^{2\eps} \!\!\!\! }{ \pi^{2\omega} \Gamma(2\eps) }\,\cdot\,
   {\rm div\!.part} \!\int\! R  \\
\hline\hline
16\eps\pslash+ & 4\eps\pslash+ & \multicolumn{2}{c|}{\ } \\
[-14pt] && \ 1\ & \mathsf qQy\ k & \frac1\eps + 3 - 2b \\
[-14pt] 16(\eps-1)m & 4(\eps-1)m & \multicolumn{2}{c|}{\ } \\
\hline
8(1-\eps)\pslash+ & 8\pslash+ & \multicolumn{2}{c|}{\ } \\
[-14pt] && n\tdot k ( n\tdot q - n\tdot k ) & \mathsf nnqQy\ k &
        \frac1{6\eps}+\frac49-\frac13b \\
[-14pt] 8(1-\eps)m & 8m & \multicolumn{2}{c|}{\ } \\
\hline
&& \multicolumn{2}{c|}{\ } & ( \frac1{12\eps} + \frac{11}{36} )
   ( n\dotg^*_p - \pslash ) \\ [-13pt]
16-16\eps & 16 & p\tdot k ( k\dotg n\tdot q - q\dotg n\tdot k ) &
   \mathsf nqqQy\ k & \\ [-12pt]
&& \multicolumn{2}{c|}{\ } & \quad +\ \frac16 b\pslash +
   \frac16 h\dotg \\
\hline                && \multicolumn{2}{c|}{\ } & \\ [-18pt]
8-8\eps & 8 & \multicolumn{2}{c|}{ \displaystyle \frac{ (m^2+p^2)
   n\tdot k ( q\dotg n\tdot k - k\dotg n\tdot q ) }
   { \mathsf nnqqQy\ k }} & \frac13 un\dotg  \\ [8pt]
\hline\hline
\end{array}  \]  \end{table}

\begin{table} \[
\begin{array} { |@{\,} c @{\,}|@{\,} c @{\,}|@{\,} c @{\,}| r @{/} l
                | l } I_{\rm c} & I_{\rm d} & I_{\rm e} &
\multicolumn{2}{c|}{{\rm integrand},\ R} & \displaystyle \frac{
   (m^2+p^2)^{2\eps} \!\!\!\! }{ \pi^{2\omega} \Gamma(2\eps) }\,\cdot\,
   {\rm div\!.part} \!\int\! R \,\dd^{2\omega} q \,\dd^{2\omega} k  \\
\hline\hline
& 16-16\eps && mn\tdot k & \mathsf nq\ yKk & m \\
-8+4\eps &&& mn\tdot k & \mathsf n\ QyKk & m \\
\hline
-4+12\eps & 8-16\eps & 4-4\eps & q\dotg\ & \mathsf qQy\ k &
   ( \frac1{2\eps} + \frac14 - c )\pslash  \\
& -8+8\eps & -8 & k\dotg\ & \mathsf qQy\ k &
   ( \frac1{4\eps} + \frac18 - \frac12c )\pslash  \\
\hline
4 & -4 && n\dotg\ \ & \mathsf n\ \ yKk &
   ( \frac1\eps - 2c ) n\dotg_p \\
2 & 4 & 2 & n\dotg\ \ & \mathsf n\ Qy\ k &
   ( \frac2\eps + 2 ) n\dotg_p - 4t n\dotg \\
-2-4\eps & 4 && n\dotg\ \ & \mathsf \ Qy\ ko & 2n\dotg_p \\
\hline
& 8 && n\dotg n\tdot p & \mathsf nqQy\ \ o & -4 ( 1 + b ) n\dotg_p +
   4(t+u) n\dotg \\
\hline
& -8 && n\dotg p\tdot q & \mathsf nq\ yKk &
   ( \frac1\eps + 3 - 2b ) n\dotg_p \\
4 &&& n\dotg p\tdot q & \mathsf n\ QyKk &
   ( \frac1\eps + 3 - 2b ) n\dotg_p \\
\hline
12-12\eps & -16+16\eps && k\dotg n\tdot k & \mathsf nq\ yKk &
   \frac38 \pslash + \frac14 n\dotg_p + \frac14 n\dotg^*_p \\
4-4\eps &&& k\dotg n\tdot k & \mathsf n\ QyKk & \frac58 \pslash +
   ( \frac1{2\eps}+\frac34 ) n\dotg_p - tn\dotg + \frac14 n\dotg^*_p \\
& 8-8\eps & 8 & k\dotg n\tdot k & \mathsf nqQy\ k & ( \frac1{6\eps}
   + \frac19 - \frac13c ) \pslash - ( \frac1{6\eps} + \frac5{18} )
   n\dotg_p + \frac13 tn\dotg \\
\hline
&&& \multicolumn{2}{c|}{\ } & \\ [-18pt]
-4+4\eps &&& \multicolumn{2}{c}{ \displaystyle \frac{ (m^2+p^2)
   k\dotg n\tdot k }{ \mathsf nqQyKk } } & un\dotg \\ [-18pt]
&&& \multicolumn{2}{c|}{\ } & \\
\hline
&&& \multicolumn{2}{c|}{\ } & \\ [-18pt]
& -4 & -2 & \multicolumn{2}{c}{ \displaystyle \frac{ (m^2+p^2) n\dotg }
   { \mathsf nqQy\ k } } & 4un\dotg \\ [-18pt]
&&& \multicolumn{2}{c|}{\ } & \\
\hline
12-12\eps &&& q\dotg n\tdot k & \mathsf nqQy\ k &
   ( \frac1{4\eps} + \frac18 - \frac12 c ) \pslash \\
-8 & 8 && q\dotg n\tdot k & \mathsf nq\ yKk &
   \frac14 \pslash + ( \frac1{2\eps} - c ) n\dotg^*_p \\
\hline
& 16 && k\dotg n\tdot p & \mathsf nq\ yKk & n\dotg^*_p \\
-12 &&& k\dotg n\tdot p & \mathsf n\ QyKk & n\dotg_p^* \\
\hline
& -8 && q\dotg n\tdot p & \mathsf nq\ yKk &
   ( \frac1\eps + 3 - 2b ) n\dotg^*_p \\
4 &&& q\dotg n\tdot p & \mathsf n\ QyKk &
   ( \frac1\eps + 3 - 2b ) n\dotg_p^* \\
& -8 & -4 & q\dotg n\tdot p & \mathsf nqQy\ k &
   2h\dotg + ( \frac1\eps + 3 ) n\dotg^*_p \\
\hline\hline
\end{array}   \]  \end{table}


\clearpage

\begin{thebibliography}{99}

\bibitem{L1} G. Leibbrandt, \RMP\,{\bf 59}, No.4 (1987) 1067.
\bibitem{L5} G. Leibbrandt, \PRD {\bf 29}\ (1984) 1699.
\bibitem{L3} G. Leibbrandt and S.-L. Nyeo, {\it Phys. Lett.} \,B {\bf
                   140}\ (1984) 417.
\bibitem{me} G. Leibbrandt and J. Williams, \NPB {\bf 440}\ (1995) 573.
\bibitem{me2} J. Williams, ``Overlapping Two-Loop Quark Self-Energy in
                   the Light-Cone Gauge'' (unpublished M.Sc. thesis,
                   University of Guelph, 1994);
\bibitem{noncovt}  A. Smith, \NPB {\bf 267}\ (1986) 277;
  \newline   G. Leibbrandt, \NPB {\bf 337}\ (1990) 87.
\bibitem{HB} G. Heinrich and Z. Kunszt, \NPB {\bf 519}\ (1998) 405
                   (hep-ph/9708334);
 \newline    A. Bassetto, G. Heinrich, Z. Kunszt, and W. Vogelsang,
                   \PRD {\bf 58} (1998) 094020 (hep-ph/9805283).
\bibitem{tH} G. 't Hooft, \NPB {\bf 33}\ (1971) 173; {\bf 35}\ (1971)
                   167.
\bibitem{co} J. C. Collins, {\em Renormalization} (Cambridge University
                   Press, Cambridge, 1984).
\bibitem{B2} A. Bassetto, M. Dalbosco, and R. Soldati, \PRD {\bf 36}
                   (1987) 3138.
\bibitem{B1} A. Bassetto, G. Nardelli, and R. Soldati, {\em Yang-Mills
                   Theories in Algebraic Non-covariant Gauges},
                   (World Scientific, Singapore, 1991).
\bibitem{L2} G. Leibbrandt, {\em Noncovariant Gauges}
                   (World Scientific, Singapore, 1994).
\bibitem{su} S.-L. Nyeo, ``Renormalization in the Light-Cone Gauge''
               (unpublished Ph.D. thesis, University of Guelph, 1986).
\bibitem{fr} J. Frenkel, \PRD {\bf 13}\ (1976) 2325.
\bibitem{ry} L. H. Ryder, {\em Quantum Field Theory} (Cambridge
                   University Press, Cambridge, 1986).
\bibitem{ma} S. Mandelstam, \NPB {\bf 213}\ (1983) 149.
\bibitem{BDLS} A. Bassetto, M. Dalbosco, I. Lazzizzera, and R. Soldati,
                   \PRD {\bf 31} (1985) 2012.
\bibitem{L6} D. M. Capper and G. Leibbrandt, \JMP\,{\bf 15}\ (1974)
                   82; 86.
\bibitem{L4} G. Leibbrandt, \RMP\,{\bf 47}\ (1975) 849.
\bibitem{we} S. Weinberg, {\em Phys. Rev.}\,{\bf 118}\ (1960) 838.
\bibitem{im} E. Mendels, {\em Nuovo Cimento} A{\bf 45}\ (1978) 87;
  \newline   A. E. Terrano, {\em Phys. Lett.} B{\bf 93}\ (1980) 424;
  \newline   G. Leibbrandt and S.-L. Nyeo, \JMP\,{\bf 27}\ (1986) 627;
  \newline   A. Bassetto, I. A. Korchemskaya, G. P. Korchemsky, and
                G. Nardelli, \NPB {\bf 408}\ (1993) 62;
  \newline   F. V. Tkachov, {\em Nucl. Instrum. Meth.} A{\bf 389}
                (1997) 309 (hep-ph/9609429);
  \newline   A. I. Davydychev, {\em Acta Phys. Polon.} B{\bf 28}
                (1997) 841 (hep-ph/9610510), and references therein;
  \newline   F. V. Tkachov, {\em Sov. J. Part. Nucl.}\,{\bf 25}\ (1994)
                649 (hep-ph/9701272);
  \newline   J. Fleischer, M. Tentyukov, and O. L. Veretin, {\em Acta
                Phys. Polon.} B{\bf 28}\ (1997) 2333 (hep-ph/9711437);
  \newline   A. I. Davydychev and P. Osland, \PRD {\bf 59}\ (1999)
                014006 (hep-ph/ 9806522);
  \newline   A. T. Suzuki and A. G. M. Schmidt, \NPB {\bf 537}\ (1999)
                549 (hep-th/ 9807158);
  \newline   N. J. Watson, \NPB {\bf 552}\ (1999) 461 (hep-ph/9812202);
  \newline   J. Fleischer and O. L. Veretin, (hep-ph/9901402).
\bibitem{TF} R. Tarrach, \NPB {\bf 183}\ (1981) 384;
  \newline   J. Fleischer, F. Jegerlehner, O. V. Tarasov, and O. L.
                Veretin, (hep-ph/ 9803493).
\bibitem{sm} N. Marcus and A. Sagnotti, \NPB {\bf 256}\ (1985) 77.
\bibitem{CB} W. E. Caswell and A. D. Kennedy, \PRD {\bf 25}\ (1982)
                392;
  \newline   G. Bonneau, {\em J. Mod. Phys.} A{\bf 5}, No.20 (1990)
                3831.
\bibitem{BX} A. Bassetto, private communication (1999).
\bibitem{VP} J. J. van der Bij and F. Hoogeveen, \NPB {\bf 283}
                (1987) 477;
  \newline   M. Peter, \NPB {\bf 501}\ (1997) 471 (hep-ph/9702245).

\end{thebibliography}
\end{document}